\documentclass[11pt]{article}
\usepackage[preprint]{acl}
\usepackage{times}
\usepackage{latexsym}
\usepackage[T1]{fontenc}
\usepackage[utf8]{inputenc}
\usepackage{microtype}
\usepackage{inconsolata}
\usepackage{booktabs}       
\usepackage{amsfonts}       
\usepackage{nicefrac}       
\usepackage{microtype}      
\usepackage{xcolor}         
\usepackage{amsmath,amssymb}
\usepackage{algorithm}

\usepackage{multirow}

\usepackage{colortbl}

\usepackage{array}
\usepackage{tabularx}

\usepackage{makecell}

\usepackage{caption}

\usepackage{rotating}

\usepackage{graphicx}

\usepackage{listings}
\usepackage{upquote}
\usepackage[most]{tcolorbox}
\tcbuselibrary{breakable,listings,skins}
\usepackage{subcaption}

\usepackage{cleveref}
\usepackage{enumitem}
\usepackage{subcaption}
\definecolor{hdr1}{RGB}{187,214,241}      
\definecolor{hdr2}{RGB}{205,226,246}      
\definecolor{hdr3}{RGB}{231,243,253}      
\definecolor{rowA}{RGB}{255,255,255}      
\definecolor{rowB}{RGB}{255,255,255}      
\definecolor{best}{RGB}{191,222,247}      
\definecolor{second}{RGB}{232,244,253}    
\definecolor{bm25col}{RGB}{217,236,250}   
\definecolor{avgrow}{RGB}{221,236,249}    
\definecolor{querycolor}{RGB}{191,222,247}
\definecolor{queryframe}{RGB}{125,164,205}
\definecolor{codecolor}{RGB}{240,247,254}
\definecolor{codeframe}{RGB}{140,176,214}
\definecolor{rewritecolor}{RGB}{247,251,255}
\definecolor{rewriteframe}{RGB}{150,184,219}
\definecolor{notecolor}{RGB}{247,250,253}
\definecolor{noteframe}{RGB}{155,171,190}
\definecolor{promptbg}{RGB}{248,248,248}
\definecolor{promptframe}{RGB}{185,185,185}
\definecolor{prompttitle}{RGB}{224,224,224}

\newcolumntype{C}[1]{>{\centering\arraybackslash}p{#1}}
\newcolumntype{L}[1]{>{\raggedright\arraybackslash}p{#1}}
\newcolumntype{Y}{>{\raggedright\arraybackslash}X}
\newcommand{\benchname}{CORE-Bench}
\newcommand{\affilmark}[1]{\textsuperscript{\mdseries #1}}
\newcommand{\bst}[1]{\cellcolor{best}\textbf{#1}}
\newcommand{\snd}[1]{\cellcolor{second}\underline{#1}}

\setlist[itemize]{leftmargin=*} 

\lstdefinestyle{wrappedtext}{
  basicstyle=\ttfamily\scriptsize,
  breaklines=true,
  breakatwhitespace=false,
  columns=fullflexible,
  keepspaces=true,
  showstringspaces=false,
  upquote=true,
  postbreak=\mbox{\textcolor{gray}{$\hookrightarrow$}\space}
}
\lstdefinestyle{promptlisting}{
  basicstyle=\ttfamily\scriptsize,
  breaklines=true,
  breakatwhitespace=false,
  columns=fullflexible,
  keepspaces=true,
  showstringspaces=false,
  upquote=true,
  postbreak=\mbox{\textcolor{gray}{$\hookrightarrow$}\space}
}
\lstdefinestyle{wrappedcode}{
  basicstyle=\ttfamily\scriptsize,
  breaklines=true,
  breakatwhitespace=false,
  columns=fullflexible,
  keepspaces=true,
  showstringspaces=false,
  upquote=true,
  keywordstyle=\color{blue!65!black}\bfseries,
  commentstyle=\color{gray}\itshape,
  stringstyle=\color{orange!70!black},
  postbreak=\mbox{\textcolor{gray}{$\hookrightarrow$}\space}
}
\lstdefinestyle{boxedcode}{
  style=wrappedcode,
  backgroundcolor=\color{codecolor},
  frame=single,
  rulecolor=\color{codeframe},
  framesep=3pt,
  linewidth=\linewidth,
  aboveskip=5pt,
  belowskip=5pt,
  captionpos=t
}
\tcbset{
  paperbox/.style={
    enhanced,
    breakable,
    width=\linewidth,
    boxsep=1pt,
    arc=1mm,
    boxrule=0.55pt,
    left=2pt,
    right=2pt,
    top=3pt,
    bottom=3pt,
    before skip=5pt,
    after skip=5pt,
    fonttitle=\bfseries\footnotesize,
    coltitle=white,
    listing only,
  },
  querybox/.style={
    paperbox,
    colback=querycolor,
    colframe=queryframe,
    colbacktitle=queryframe,
    listing options={style=wrappedtext}
  },
  codebox/.style={
    paperbox,
    colback=codecolor,
    colframe=codeframe,
    colbacktitle=codeframe,
    listing options={style=wrappedcode}
  },
  rewritebox/.style={
    paperbox,
    colback=rewritecolor,
    colframe=rewriteframe,
    colbacktitle=rewriteframe,
    listing options={style=wrappedtext}
  },
  querytext/.style={
    enhanced,
    width=\linewidth,
    boxsep=1pt,
    arc=1mm,
    boxrule=0.55pt,
    left=2pt,
    right=2pt,
    top=3pt,
    bottom=3pt,
    before skip=5pt,
    after skip=5pt,
    colback=querycolor,
    colframe=queryframe,
    colbacktitle=queryframe,
    coltitle=white,
    fonttitle=\bfseries\footnotesize,
    fontupper=\ttfamily\scriptsize
  },
  rewritetext/.style={
    enhanced,
    width=\linewidth,
    boxsep=1pt,
    arc=1mm,
    boxrule=0.55pt,
    left=2pt,
    right=2pt,
    top=3pt,
    bottom=3pt,
    before skip=5pt,
    after skip=5pt,
    colback=rewritecolor,
    colframe=rewriteframe,
    colbacktitle=rewriteframe,
    coltitle=white,
    fonttitle=\bfseries\footnotesize,
    fontupper=\ttfamily\scriptsize
  },
  compactnote/.style={
    enhanced,
    breakable,
    width=\linewidth,
    boxsep=1pt,
    arc=1mm,
    boxrule=0.55pt,
    left=3pt,
    right=3pt,
    top=3pt,
    bottom=3pt,
    before skip=5pt,
    after skip=5pt,
    colback=notecolor,
    colframe=noteframe,
    colbacktitle=noteframe,
    coltitle=white,
    fonttitle=\bfseries\footnotesize,
  },
  promptbox/.style={
    enhanced,
    width=\textwidth,
    boxsep=1pt,
    arc=1mm,
    boxrule=0.55pt,
    left=4pt,
    right=4pt,
    top=4pt,
    bottom=4pt,
    before skip=6pt,
    after skip=6pt,
    colback=promptbg,
    colframe=promptframe,
    colbacktitle=prompttitle,
    coltitle=black,
    fonttitle=\bfseries\footnotesize,
    listing only,
    listing options={style=promptlisting}
  }
}
\title{\benchname{}: A Comprehensive Benchmark for Code Retrieval in the Era of Agentic Coding}

\author{
  \textbf{Fuwei Zhang}\affilmark{1} \quad
  \textbf{Yanzhao Zhang} \quad
  \textbf{Mingxin Li} \quad
  \textbf{Dingkun Long}
  \\
  \textbf{Lexiang Hu}\affilmark{2} \quad
  \textbf{Pengjun Xie} \quad
  \textbf{Zhao Zhang}\affilmark{3} \quad
  \textbf{Fuzhen Zhuang}\affilmark{1}
  \\
  \affilmark{1}Institute of Artificial Intelligence, Beihang University
  \\
  \affilmark{2}State Key Lab of General AI, School of Intelligence Science and Technology, Peking University
  \\
  \affilmark{3}SKLCCSE, School of Computer Science and Engineering, Beihang University, China
}

\begin{document}
\maketitle
\begin{abstract}
Code retrieval is becoming central to coding agents, but agentic coding requires more than matching a natural-language query to an isolated snippet. Given a user request, a coding agent needs to navigate a concrete repository state, locate relevant files and functions, gather supporting context, and filter similar in-repository distractors. Existing code retrieval benchmarks mainly evaluate docstring-to-function or snippet-level matching, thereby missing this requirement-driven repository search problem. To address this gap, we introduce \textbf{\benchname{}}\footnotemark, a comprehensive benchmark for code retrieval in the era of agentic coding. \benchname{} evaluates code retrieval ability at three levels: code understanding, issue-to-edit localization, and broader context retrieval. Built from curated code-search tasks and SWE-bench-series instances, \benchname{} contains over 180K queries and 106K broader-context relevance labels. Experiments with representative embedding models show a sharp drop from traditional code search to code retrieval in agentic coding settings. Simple supervised fine-tuning of existing embedding models significantly improves performance in this setting, suggesting substantial room for further progress.
\end{abstract}
\footnotetext{The dataset is available on \href{https://huggingface.co/datasets/zhangfw123/CORE-Bench}{Hugging Face}, and the evaluation code is available on \href{https://github.com/zhangfw123/CORE-Bench-Eval}{GitHub}.}

\section{Introduction}

Code retrieval has long served as a bridge between natural language and code. In agentic coding and vibe-coding settings, retrieval becomes part of the coding-agent workflow: given a bug report, feature request, refactoring request, or other user request, an agent must decide which files, functions, and documents are worth inspecting before editing. Current coding-agent harnesses, such as OpenHands~\cite{wang2025openhands} and SWE-agent~\cite{yang2024swe}, usually do not reduce these decisions to a single embedding lookup. Instead, they rely heavily on command-line repository exploration, including \texttt{ls}, \texttt{grep}, \texttt{find}, file reading, and iterative narrowing. These steps show that retrieval remains central in practice, but the retrieval problem shifts toward interactive, requirement-driven repository search. This raises a concrete question: why are retrieval models that perform well on standard code-search benchmarks often less useful for this search process?
This mismatch suggests that code retrieval in agentic coding settings may be better understood as \emph{requirement-driven repository search}, rather than isolated code-snippet matching. Existing benchmarks are not fully aligned with this setting. CodeSearchNet~\cite{husain2019codesearchnet} and many code tasks in MTEB~\cite{muennighoff2022mteb} and CoIR~\cite{li2024coir} mainly evaluate decontextualized snippet retrieval under fixed corpora and simplified query-document mappings. These benchmarks measure basic natural-language/code alignment but miss several properties that matter for code retrieval in agentic coding: (1) realistic development requests often have a large intent-to-implementation gap; (2) relevant evidence is heterogeneous and scattered across code, configuration files, dependencies, and other repository artifacts; (3) candidate documents are often long function-level chunks or documentation snippets, which can dilute the evidence captured by a single embedding; (4) a single request may require multiple edit snippets and related context; and (5) real repositories contain dense local distractors, such as similar wrappers, adapters, and configuration snippets. Strong performance on existing benchmarks may therefore overstate a model's usefulness for coding agents.
To address this gap, we introduce \textbf{\benchname{}}, a comprehensive code retrieval benchmark for evaluating retrieval ability in agentic coding settings. \textbf{\benchname{}} has three levels. \textbf{Level-1: Code Understanding} retains challenging traditional retrieval tasks to evaluate foundational code-understanding ability. \textbf{Level-2: Issue-to-Edit Localization} tests whether a model can retrieve the files or code chunks that must be modified for a requested change. \textbf{Level-3: Broader Context Retrieval} further evaluates whether a model can retrieve auxiliary code, documents, and other information needed to understand and complete the change.
To align with existing agentic coding evaluation settings, \benchname{} further annotates datasets from the SWE-bench series. Patch information provides direct labels for issue-to-edit localization, while the corresponding repository snapshots preserve the pre-resolution code state and simulate the actual programming environment. For broader context retrieval, we build an automated pipeline that combines coarse annotation, relevance filtering, and quality validation to identify additional context useful for completing the requested change. The resulting data contain realistic requirements, temporally grounded repository corpora, and multi targets over edited code and broader context.
We evaluate general and code-oriented embedding models of different scales on \textbf{\benchname{}}. In addition, in-domain supervised fine-tuning~(SFT) with supervision constructed from pull requests in open-source GitHub repositories substantially improves embedding models. 
Here are our contributions:
\begin{itemize}[leftmargin=*,topsep=2pt,itemsep=0pt,parsep=0pt]
    \item We find existing code retrieval methods fall short in agentic coding, and revisit code retrieval for agentic coding as requirement-driven repository search and summarize key challenges missed by existing benchmarks.
    \item We introduce \textbf{\benchname{}}, a three-level benchmark for evaluating sparse and dense retrievers in requirement-driven repository search for agentic coding, covering code understanding, issue-to-edit localization, and broader context retrieval.
    \item We show that current embedding retrievers suffer large performance drops on code retrieval in agentic coding settings, while in-domain SFT with supervision from GitHub pull requests improves performance but still leaves a clear gap.
\end{itemize}

\section{Related Work}

\subsection{Code Search Benchmarks}

Existing code retrieval benchmarks mostly ask a query to retrieve an ``answer'' artifact, rather than agent-useful repository context. CodeSearchNet \cite{husain2019codesearchnet}, CodeXGLUE \cite{lu1codexglue}, CoSQA \cite{huang2021cosqa}, APPS \cite{hendrycks2021apps}, HumanEval \cite{chen2021humaneval,muennighoff2022mteb}, MBPP \cite{austin2021mbpp,muennighoff2022mteb}, and DS-1000 \cite{lai2022ds1000,muennighoff2022mteb} cover docstrings, web queries, programming problems, algorithms, and data science. CodeTransOcean \cite{yan2023codetransocean}, CodeEditSearch \cite{muennighoff2022mteb}, StackOverflowQA \cite{muennighoff2022mteb}, FreshStack \cite{thakur2025freshstack,muennighoff2022mteb}, and CodeRAG-Bench \cite{wang2025coderagbench} broaden the surface to multilingual code-to-code retrieval, version-history edits, developer text, mixed code--natural-language artifacts, and repositories with external knowledge sources.

Umbrella benchmarks such as MTEB \cite{muennighoff2022mteb}, CoIR \cite{li2024coir}, CLARC \cite{wang2026clarc}, and CPRet \cite{deng2025cpret} unify task formats and domains. SWELoc \cite{SweRankEmbed-Large}, LOC-BENCH \cite{chen2025locagent}, and ContextBench \cite{li2026contextbench} approach agentic coding, but mainly study process analysis or edit-location finding, leaving standalone retrievers for multi-location context retrieval under concrete repository states and local distractors underexplored.

\begin{figure*}[t]
\centering
\begin{subfigure}[t]{0.48\textwidth}
    \centering
    \includegraphics[width=\linewidth]{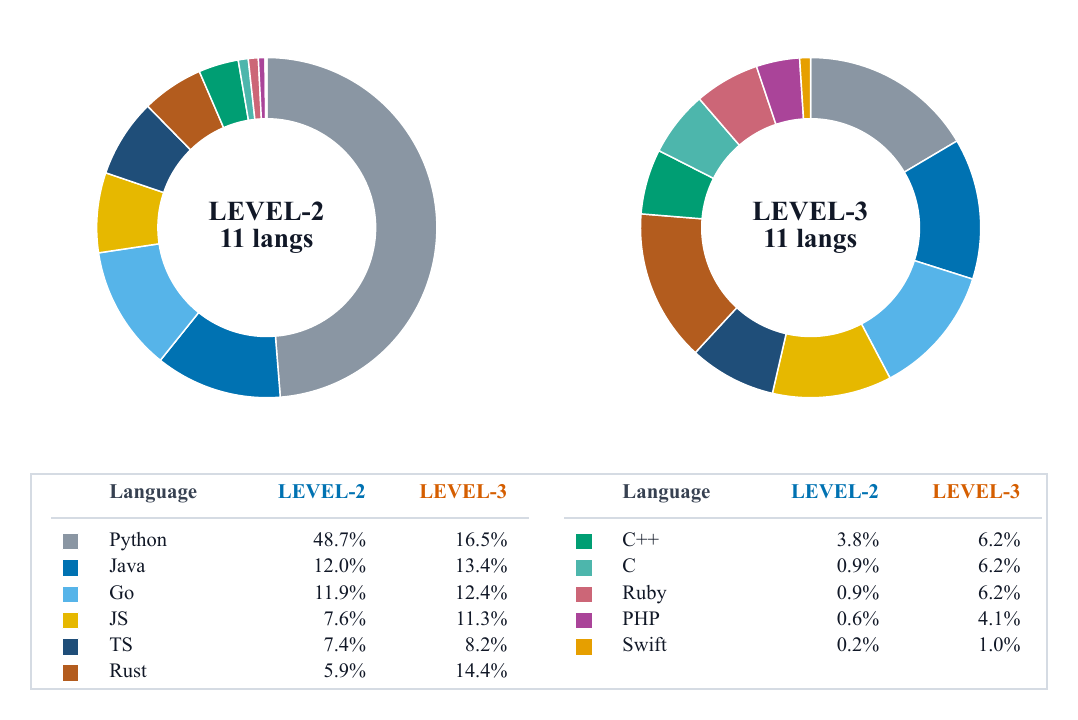}
    \caption{Repository language coverage.}
    \label{fig:dataset-overview-language}
\end{subfigure}
\hfill
\begin{subfigure}[t]{0.48\textwidth}
    \centering
    \includegraphics[width=\linewidth]{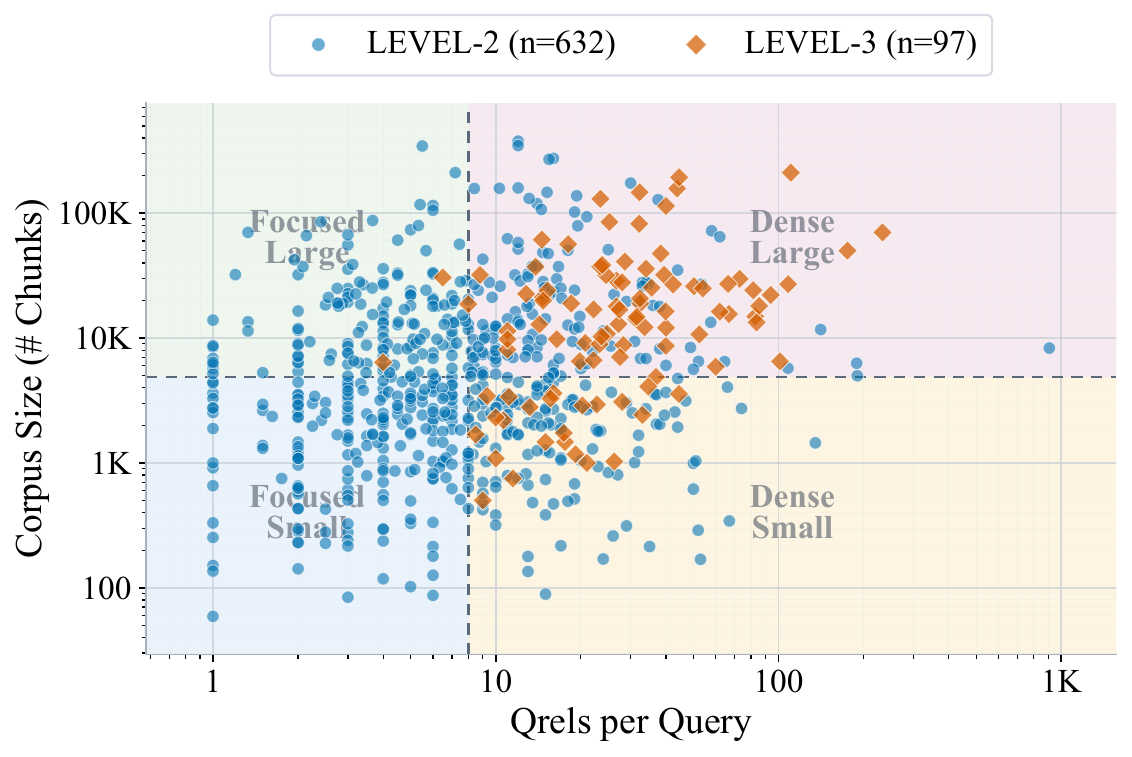}
    \caption{Repository-level retrieval difficulty.}
    \label{fig:dataset-overview-difficulty}
\end{subfigure}
\caption{Analysis of \benchname{} (Level-2 and Level-3), including language coverage and difficulty.}
\label{fig:dataset-overview-analysis}
\end{figure*}

\begin{table*}[t]
\centering
\caption{Statistics of \benchname{}. Rels/Q denotes number of relevance documents per query. Avg.Q/Avg.C denote mean query/corpus lens.}
\label{tab:dataset_stats}
\setlength{\tabcolsep}{4.2pt}
\small
\renewcommand{\arraystretch}{1.15}
\resizebox{\textwidth}{!}{%
\begin{tabular}{llrrrrrrr}
\toprule
\textbf{Setting}
 & \textbf{Task}
 & \textbf{\#Repos}
 & \textbf{\#Queries}
 & \textbf{\#Corpus}
 & \textbf{\#Qrels}
 & \textbf{Rels/Q}
 & \textbf{Avg.Q Len}
 & \textbf{Avg.C Len}
\\
\midrule
\rowcolor{rowA}
Level-1 & Code Understanding & -- & 172{,}961 & 2{,}412{,}296 & 823{,}345 & 4.8 & 676 & 412 \\
\rowcolor{rowB}
Level-2 & Issue-to-Edit Localization & 632 & 5{,}061 & 9{,}377{,}120 & 52{,}712 & 10.42 & 1{,}558.5 & 1{,}004.9 \\
\rowcolor{rowA}
Level-3 & Broader Context Retrieval & 97 & 2{,}580 & 2{,}609{,}581 & 106{,}479 & 41.27 & 1{,}074.4 & 1{,}154.8 \\
\bottomrule
\end{tabular}}
\vspace{2pt}
\end{table*}

\begin{table}[t]
\centering
\caption{Query intent distribution for Level-2/3.}
\label{tab:query_type_distribution}
\setlength{\tabcolsep}{4.2pt}
\small
\renewcommand{\arraystretch}{1.15}
\begin{tabular}{lrr}
\toprule
\textbf{Intent} & \textbf{Level-2} & \textbf{Level-3} \\
\midrule
Bug report & 3,047 (60.2\%) & 1,468 (56.9\%) \\
Feature request & 1,659 (32.8\%) & 927 (35.9\%) \\
Refactoring & 187 (3.7\%) & 121 (4.7\%) \\
Question & 91 (1.8\%) & 11 (0.4\%) \\
Other & 59 (1.2\%) & 44 (1.7\%) \\
Documentation & 18 (0.4\%) & 9 (0.3\%) \\
\midrule
\textbf{Total} & \textbf{5,061} & \textbf{2,580} \\
\bottomrule
\end{tabular}
\end{table}

\subsection{Code Search Models}

Code search models span lexical, sparse, dense, late-interaction, and hybrid retrieval. BM25 \cite{BM25}, bag-of-words analyses \cite{zhang2021bowcodesearch}, and SPLADE-style sparse expansion \cite{formal2021spladev2} remain useful for exact identifiers, APIs, file names, stack traces, and configuration keys.

Dense methods such as DPR \cite{karpukhin2020dpr} and ANCE \cite{xiong2021ance} introduced dual encoders and hard-negative mining; general embedding families multilingual-E5 \cite{multilingual-e5}, BGE \cite{bge-large-en-v1.5}, BGE-M3 \cite{bge-m3}, GTE \cite{gte-multilingual-base}, GTE-Qwen \cite{gte-Qwen2-1.5B-instruct}, Qwen3-Embedding \cite{Qwen3-Embedding}, E5-Mistral \cite{e5-mistral-7b-instruct}, pplx-embed \cite{pplx-embed-v1-4b}, Jina Embeddings \cite{jina-embeddings-v4,jina-embeddings-v5-text-small}, and F2LLM-v2 \cite{F2LLM-v2} provide strong transfer baselines. Code-oriented models CodeBERT \cite{feng2020codebert}, GraphCodeBERT \cite{guo2021graphcodebert}, UniXcoder \cite{guo2022unixcoder}, CodeXEmbed \cite{liu2025codexembed}, CodeRankEmbed \cite{CodeRankEmbed}, Jina Code Embeddings \cite{jina-code-embeddings}, C2LLM \cite{C2LLM-7B}, and SweRankEmbed \cite{SweRankEmbed-Large} add programming-language or issue-localization supervision, while ColBERT \cite{khattab2020colbert} and BEIR \cite{thakur2021beir} motivate late-interaction or hybrid matching. Yet these models are not designed for requirement-driven, repository-state-aware code-agent retrieval.

Overall, modern agents such as OpenHands \cite{wang2025openhands} and SWE-agent \cite{yang2024swe} need retrieval over concrete repository states, multiple edit- and reasoning-relevant locations, and plausible in-repository distractors. \benchname{} targets this missing code-agent retrieval setting.

\section{Benchmark}
\begin{figure*}[t]
      \centering
      \includegraphics[width=\linewidth]{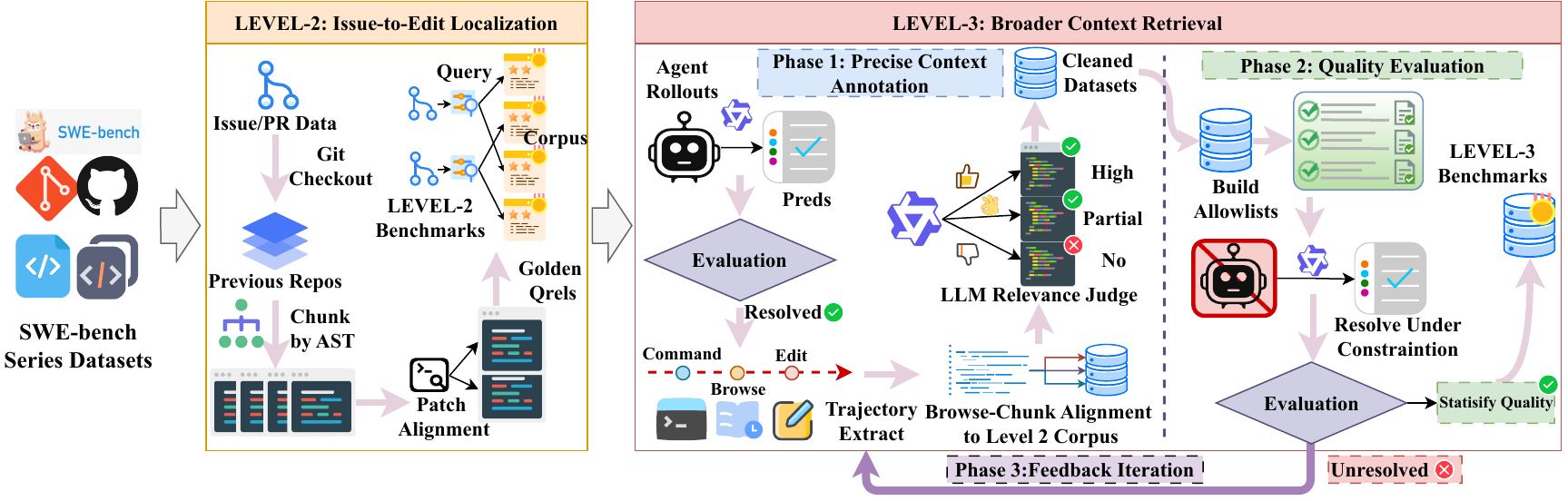}
      \caption{Overall pipeline of \benchname{} (LEVEL-2 and LEVEL-3) benchmark construction. }
      \label{fig:pipeline}
\end{figure*}

\subsection{Overview of \benchname{}}

\benchname{} uses a three-level evaluation:
\begin{itemize}[leftmargin=*,topsep=2pt,itemsep=0pt,parsep=0pt]
    \item \textbf{Level-1: Code Understanding.} We retain challenging traditional retrieval tasks across Code-to-Text, Text-to-Code, Code-to-Code, and hybrid settings, excluding datasets with weak alignment with realistic retrieval.
    \item \textbf{Level-2: Issue-to-Edit Localization.} We evaluate whether a model can retrieve files or chunks that must be modified for a requested change, using patch-aligned edit snippets from SWE-bench-series and SWE-bench-style software-change datasets.
    \item \textbf{Level-3: Broader Context Retrieval.} We evaluate whether retrieval can recover auxiliary code, documentation, and tests that help an agent reason about and complete the change, using an automated annotation pipeline because such context is rarely recorded by developers.
\end{itemize}

Table~\ref{tab:dataset_stats} summarizes the aggregate scale of \benchname{}.
Level-1 contains 172K queries over 2.4M corpus items for foundational code
understanding. Level-2 covers 632 repositories, 5{,}061 queries, and 9.38M
repository chunks for issue-to-edit localization, while Level-3 contains 2{,}580
queries and 106K relevance labels for broader context retrieval. Detailed
sub-dataset statistics are provided in Appendix~\ref{app:dataset-detail-stats}.
Although Level-3 has fewer queries than Level-2, it contains roughly twice as
many qrels and nearly four times as many relevant chunks per query, reflecting
its denser broader-context annotation target.
We also provide rewrite variants for Level-2 and Level-3 by converting raw
pull request (PR) or issue descriptions into concise developer-facing queries
that resemble what users would ask an AI coding assistant, while keeping the
corpus and qrels unchanged.

Figure~\ref{fig:dataset-overview-analysis} and Table~\ref{tab:query_type_distribution}
further characterize the agentic levels. Across Level-2 and Level-3,
\benchname{} spans 11 programming languages, as shown in
Figure~\ref{fig:dataset-overview-analysis}(\subref{fig:dataset-overview-language}); bug reports and feature requests
account for over 92\% of queries. Figure~\ref{fig:dataset-overview-analysis}(\subref{fig:dataset-overview-difficulty})
groups repositories by retrieval corpus size and relevant chunks per query,
yielding four regimes that test both precise edit localization and large-scale
contextual retrieval.

\subsection{Benchmark Construction}

\subsubsection{LEVEL-1: Code Understanding}

For Level-1, we curate existing code-retrieval datasets into a unified retrieval setting \cite{li2024coir,muennighoff2022mteb}. The retained benchmark covers Text-to-Code, Code-to-Text, Code-to-Code, hybrid retrieval, and competitive-programming retrieval through APPS, CoSQA, SyntheticText2SQL, CLARC, CodeSearchNet, CodeTransOcean-DL, CodeFeedback-ST, and CPRet \cite{hendrycks2021apps,huang2021cosqa,muennighoff2022mteb,wang2026clarc,husain2019codesearchnet,yan2023codetransocean,zheng-etal-2024-opencodeinterpreter,deng2025cpret}. We exclude saturated or weakly aligned tasks, including StackOverflowQA, CodeEditSearchRetrieval, and CodeFeedback-MT.

\subsubsection{LEVEL-2: Issue-to-Edit Localization}

LEVEL-2 is built from existing end-to-end agent benchmarks in the SWE-bench family, such as SWE-bench\_Verified, SWE-bench\_Pro, and related SWE-bench-style datasets. The LLM-based query-filtering stage uses \mbox{Qwen3.5-397B-A17B}~\cite{qwen3.5} to remove answer-leaking, low-quality, underspecified, or non-actionable PR and issue descriptions before patch alignment.

As shown in Figure~\ref{fig:pipeline}, the construction proceeds in five steps. First, we collect PR or issue metadata from SWE-bench-derived sources. Second, we checkout each repository to the commit immediately before the PR, so the retrieval corpus matches the code state seen by the corresponding request. Third, we chunk source and documentation files with AST and LangChain-based splitters, filtering irrelevant file types while preserving file paths and line spans. Fourth, we align the PR patch back to these pre-PR chunks by matching modified files and edited line ranges. Finally, we treat the PR or issue description as the query and the patch-aligned chunks as relevant documents. This turns historical software changes into issue-to-edit localization instances without using post-hoc diffs as retrievable answers.

\label{sec:methodology}

\subsubsection{LEVEL-3: Broader Context Retrieval}
\label{sec:level3-construction}
Level-3 extends Level-2 beyond edited chunks. In real development tasks, a developer or agent often reads tests, helper functions, configuration files, documentation, and neighboring call sites before deciding how to edit the code. Such supporting context is rarely preserved in PR metadata, so we construct it with an automated pipeline that reruns resolution attempts and extracts useful context from the resulting agent execution traces. The pipeline has three phases: \textbf{1) Precise Context Annotation; 2) Quality Verification; and 3) Feedback Iteration.}

\noindent \textbf{Phase 1: Precise Context Annotation} 
We combine a Code Agent with LLM-based relevance assessment. Mini-SWE-Agent reruns the corresponding development tasks with \mbox{Qwen3.5-397B-A17B}. From the execution traces, we extract browse actions, including files opened by commands such as \texttt{cat}, \texttt{grep}, \texttt{head}, and \texttt{sed}. These browsed snippets are aligned to the Level-2 chunking schema, so overlapping Level-2/3 instances use comparable chunk units. The resulting query--chunk pairs are still noisy because agents may inspect irrelevant files. We therefore apply a three-vote LLM relevance ensemble: \mbox{Qwen3.5-397B-A17B} casts two independent votes, and \mbox{Claude Sonnet 4.6}~\cite{claude-sonnet-4-6} casts one additional vote for each pair.

\noindent \textbf{Phase 2: Quality Evaluation} 
We validate annotations by modifying mini-swe-agent to run in a closed allowlist setting. For each task, the agent can only read files covered by the annotation-derived allowlist. We build one allowlist from Level-2 edit labels and another from Level-3 broader-context labels, then rerun the annotated dataset under both settings. A higher resolution rate under the Level-3 allowlist indicates that the added context is not merely textually related, but functionally useful for completing the requested change.

\noindent \textbf{Phase 3: Feedback Iteration} 
For cases whose annotations remain insufficient, we perform feedback iteration. Unresolved or weakly supported cases are rerun, and newly browsed contextual snippets are aligned, judged, and added as supplementary labels. We do not require every retained trajectory to end in a solved patch, because resolved runs tend to favor easier tasks with smaller edits. Instead, useful trajectory-derived context can be kept after relevance judging when it helps understand or localize the requested change.

\subsection{Rewrite Version of \benchname{}}
Original PR and issue descriptions are often detailed and often include issue-template boilerplate, whereas practical agentic or vibe-coding queries are usually shorter, more casual, and directly addressed to an AI coding assistant. For the retained Level-2 and Level-3 queries, we therefore provide rewritten variants that preserve key technical terms, error messages, and code clues while removing irrelevant template noise and redundant context. The separate query-filtering stage handles unsuitable inputs before benchmark construction; rewriting only changes the retained query style. Both stages use \mbox{Qwen3.5-397B-A17B}, and the prompt templates are provided in Appendix~\ref{app:prompt-templates}.

\subsection{Training Data Construction}

To evaluate whether LLM embedding models can benefit from task-specific supervision, we construct an in-domain SFT set from PRs outside the SWE-bench family. PR descriptions serve as queries, while patch-aligned chunks provide positive supervision and repository-local chunks provide hard negatives. The full data scale, sampling strategy, and contrastive training objective are provided in Appendix~\ref{app:sft-training-details}.

\begin{table*}[t]
\centering
\footnotesize
\caption{%
  \textbf{\benchname{} results on a model subset (NDCG@10/Recall@100).}
  \textbf{L1} = Level-1 Code Understanding; \textbf{L2} = Level-2 Issue-to-Edit Localization;
  \textbf{L3} = Level-3 Broader Context Retrieval; \textbf{Rw} = query-rewrite variant.
}
\label{tab:overall}
\setlength{\tabcolsep}{1.4pt}
\renewcommand{\arraystretch}{1.08}
\begin{tabular*}{\textwidth}{@{\extracolsep{\fill}}llccccccc@{}}
\toprule
\textbf{Model} & \textbf{\#Param} & \textbf{L1} & \textbf{L2} & \textbf{L3} & \textbf{L2-Rw} & \textbf{L3-Rw} & \textbf{L1--L3 Avg.} & \textbf{Rewrite Avg.} \\
\cmidrule(lr){1-9}
\rowcolor{black!4}\multicolumn{9}{c}{\small\textbf{\textit{Small Embedding Models \hspace{0.35em}(<1B)}}}\\
\cmidrule(lr){1-9}
CodeRankEmbed & 137M & 47.4/78.6 & 12.1/32.9 & 22.5/28.6 & 13.8/37.9 & 26.2/31.4 & 27.3/46.7 & 20.0/34.6 \\
bge-m3 & 568M & 23.8/53.8 & 4.6/18.3 & 8.7/12.5 & 3.9/15.3 & 7.8/11.4 & 12.4/28.2 & 5.8/13.3 \\
Qwen3-Embedding-0.6B & 0.6B & 66.9/94.7 & 17.0/45.5 & 32.6/40.2 & 16.5/45.5 & 31.1/39.2 & 38.8/60.1 & 23.8/42.3 \\
\cmidrule(lr){1-9}
\rowcolor{black!4}\multicolumn{9}{c}{\small\textbf{\textit{Medium Embedding Models \hspace{0.35em}(1B--2B)}}}\\
\cmidrule(lr){1-9}
gte-Qwen2-1.5B-instruct & 1.5B & 31.3/63.0 & 3.5/15.9 & 8.1/14.9 & 4.3/18.2 & 9.5/15.7 & 14.3/31.3 & 6.9/16.9 \\
jina-code-embeddings-1.5b & 1.5B & 56.2/88.7 & 17.0/48.5 & 31.6/42.7 & 14.9/46.4 & 28.9/39.2 & 34.9/60.0 & 21.9/42.8 \\
\cmidrule(lr){1-9}
\rowcolor{black!4}\multicolumn{9}{c}{\small\textbf{\textit{Large Embedding Models \hspace{0.35em}($\geq$4B)}}}\\
\cmidrule(lr){1-9}
Qwen3-Embedding-4B & 4B & \cellcolor{best}\textbf{72.7}/\textbf{96.9} & 18.3/46.9 & 32.8/40.8 & 17.3/46.2 & 31.1/39.3 & 41.3/61.5 & 24.2/42.8 \\
C2LLM-7B & 7B & \cellcolor{second}\underline{72.4}/96.7 & 16.7/44.9 & 32.9/41.0 & 14.5/41.4 & 28.4/37.8 & 40.7/60.9 & 21.5/39.6 \\
e5-mistral-7b-instruct & 7B & 54.7/86.4 & 18.3/51.7 & 33.9/44.4 & 17.4/49.9 & 31.8/42.0 & 35.6/60.9 & 24.6/45.9 \\
SweRankEmbed-Large & 7B & 45.4/79.1 & 22.4/52.1 & 34.4/42.4 & 21.7/49.7 & 33.6/41.5 & 34.0/57.9 & 27.7/45.6 \\
F2LLM-v2-8B & 8B & 63.8/91.9 & 19.0/45.8 & 33.8/37.2 & 17.8/43.3 & 30.8/34.9 & 38.9/58.3 & 24.3/39.1 \\
Qwen3-Embedding-8B & 8B & 71.7/\underline{96.9} & 20.3/48.0 & 34.4/41.5 & 17.2/45.1 & 30.8/38.4 & 42.2/62.1 & 24.0/41.7 \\
\cmidrule(lr){1-9}
\rowcolor{black!4}\multicolumn{9}{c}{\small\textbf{\textit{In-domain SFT Models \hspace{0.35em}(0.6B--8B)}}}\\
\cmidrule(lr){1-9}
Qwen3-0.6B-SFT & 0.6B & 58.1/90.3 & 26.5/59.4 & 44.5/54.4 & 25.5/58.6 & 40.2/52.0 & 43.0/68.1 & 32.8/55.3 \\
Qwen3-4B-SFT & 4B & 59.8/90.5 & \cellcolor{second}\underline{30.3}/\underline{66.0} & \cellcolor{second}\underline{49.2}/\textbf{61.6} & \cellcolor{second}\underline{27.3}/\textbf{64.3} & \cellcolor{second}\underline{44.1}/\textbf{58.1} & \cellcolor{second}\underline{46.4}/\underline{72.7} & \cellcolor{second}\underline{35.7}/\textbf{61.2} \\
Qwen3-8B-SFT & 8B & 63.0/92.5 & \cellcolor{best}\textbf{32.8}/\textbf{66.4} & \cellcolor{best}\textbf{50.2}/\underline{61.4} & \cellcolor{best}\textbf{29.2}/\underline{64.0} & \cellcolor{best}\textbf{44.4}/\underline{57.4} & \cellcolor{best}\textbf{48.7}/\textbf{73.4} & \cellcolor{best}\textbf{36.8}/\underline{60.7} \\
\bottomrule
\end{tabular*}
\end{table*}

\section{Experiments and Evaluation}

\subsection{Experimental Setup}
\paragraph{Baselines.}
We select a representative subset of embedding models for the main-paper comparison, including BGE-M3 \cite{bge-m3}, GTE-Qwen2-1.5B-instruct \cite{gte-Qwen2-1.5B-instruct}, Qwen3-Embedding \cite{Qwen3-Embedding}, E5-Mistral \cite{e5-mistral-7b-instruct}, F2LLM-v2 \cite{F2LLM-v2}, CodeRankEmbed \cite{CodeRankEmbed}, Jina Code Embeddings \cite{jina-code-embeddings}, C2LLM-7B \cite{C2LLM-7B}, SweRankEmbed-Large \cite{SweRankEmbed-Large}, and our in-domain fine-tuned Qwen3-Embedding variants. All in-domain SFT rows report checkpoints after three supervised fine-tuning epochs. A more complete list of evaluated baselines and their model descriptions is provided in Appendix~\ref{app:baseline-models}.

\paragraph{Evaluation Details.}
We evaluate all embedding models with a modified MTEB framework \cite{muennighoff2022mteb}. Instead of materializing one corpus for each query, we build a repository-level corpus by merging and deduplicating chunks from different temporal snapshots, while preserving repository and commit metadata for each chunk. During scoring, the evaluator filters this merged corpus separately for each query, so metrics are computed only over the documents visible at the corresponding PR or issue state. For instruction-tuned models, we follow each model family's official query and document instructions, and use raw text when no template is available. We report NDCG@10 and Recall@100, with metric definitions provided in Appendix~\ref{app:evaluation-metrics}. NDCG@10 measures top-rank precision for LEVEL-2 edit localization, while Recall@100 measures whether LEVEL-3 retrieves the broader context needed by the agent.

\subsection{Evaluation of Retrieval Models}
\begin{figure}[!t]
\centering
\includegraphics[width=0.9\columnwidth]{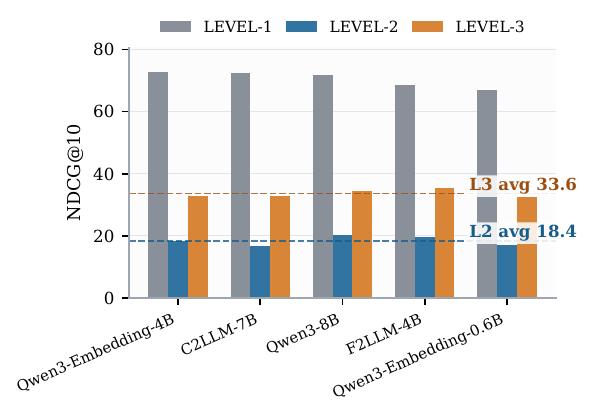}
\caption{Detailed performance comparison of top models.}
\label{fig:level1-agentic-drop}
\end{figure}

Table~\ref{tab:overall} summarizes representative results across the three \benchname{} levels and their query-rewrite variants; complete results are provided in Appendix~\ref{app:detailed-results}. We observe a large gap between traditional code understanding and repository-level retrieval for agentic coding. Many strong code-search models collapse to a similar low range on LEVEL-2/3, as shown in Figure~\ref{fig:level1-agentic-drop}. For example, Qwen3-Embedding-8B reaches 71.7/96.9 on LEVEL-1, but drops to 20.3/48.0 on LEVEL-2 and 34.4/41.5 on LEVEL-3, indicating that LEVEL-2/3 are not simply scaled-up versions of traditional code search.

LEVEL-2 has lower NDCG@10, whereas LEVEL-3 has lower Recall@100, reflecting their different focus on edit-location precision and broader context coverage. Rewritten queries often fail to improve retrieval and can even reduce performance, suggesting that realistic repository retrieval depends on request-specific context that may be lost during rewriting.

\begin{figure}[!t]
\centering
\includegraphics[width=0.9\columnwidth]{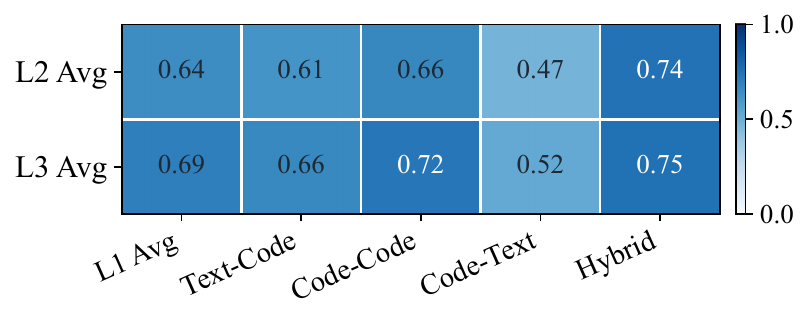}
\caption{Spearman correlation of NDCG@10 between LEVEL-1 tasks and LEVEL-2/3 averages.}
\label{fig:level-correlation}
\end{figure}

\begin{table*}[!t]
\centering
\scriptsize
\caption{%
  \textbf{\benchname{} per-language results for a subset of models on LEVEL-2 and LEVEL-3 (NDCG@10).}
}
\label{tab:language-results}
\setlength{\tabcolsep}{1pt}
\renewcommand{\arraystretch}{1.06}
\begin{tabular*}{\textwidth}{@{\extracolsep{\fill}}l*{22}{c}@{}}
\toprule
& \multicolumn{11}{c}{\textbf{LEVEL-2}} & \multicolumn{11}{c}{\textbf{LEVEL-3}} \\
\cmidrule(lr){2-12} \cmidrule(lr){13-23}
\textbf{Model} & \rotatebox{65}{\textbf{Python}} & \rotatebox{65}{\textbf{Go}} & \rotatebox{65}{\textbf{JS}} & \rotatebox{65}{\textbf{Rust}} & \rotatebox{65}{\textbf{TS}} & \rotatebox{65}{\textbf{Java}} & \rotatebox{65}{\textbf{C++}} & \rotatebox{65}{\textbf{C}} & \rotatebox{65}{\textbf{Ruby}} & \rotatebox{65}{\textbf{PHP}} & \rotatebox{65}{\textbf{Swift}} & \rotatebox{65}{\textbf{Python}} & \rotatebox{65}{\textbf{Go}} & \rotatebox{65}{\textbf{JS}} & \rotatebox{65}{\textbf{Rust}} & \rotatebox{65}{\textbf{TS}} & \rotatebox{65}{\textbf{Java}} & \rotatebox{65}{\textbf{C++}} & \rotatebox{65}{\textbf{C}} & \rotatebox{65}{\textbf{Ruby}} & \rotatebox{65}{\textbf{PHP}} & \rotatebox{65}{\textbf{Swift}} \\
\cmidrule(lr){1-23}
\rowcolor{black!4}\multicolumn{23}{c}{\textbf{\textit{Small Embedding Models \hspace{0.35em}(<1B)}}}\\
\cmidrule(lr){1-23}
CodeRankEmbed & 10.7 & 15.3 & 8.8 & 12.5 & 10.2 & 8.7 & 14.1 & 12.4 & 5.4 & 6.4 & 0.0 & 26.4 & 33.5 & 12.3 & 26.5 & 17.6 & 21.7 & 28.6 & 26.3 & 13.6 & 10.6 & 0.0 \\
bge-m3 & 5.1 & 4.5 & 4.0 & 4.7 & 2.2 & 3.0 & 3.8 & 3.0 & 2.6 & 2.1 & 0.0 & 8.1 & 10.6 & 5.0 & 10.9 & 4.1 & 8.3 & 10.0 & 8.9 & 3.7 & 2.0 & 0.0 \\
Qwen3-0.6B & 18.4 & 16.5 & 18.1 & 14.3 & 17.0 & 10.2 & 17.7 & 11.8 & 8.0 & 4.5 & 17.7 & 38.2 & 37.7 & 27.5 & 34.8 & 34.1 & 30.4 & 37.4 & 28.8 & 26.0 & 12.0 & 7.1 \\
\cmidrule(lr){1-23}
\rowcolor{black!4}\multicolumn{23}{c}{\textbf{\textit{Medium Embedding Models \hspace{0.35em}(1B--2B)}}}\\
\cmidrule(lr){1-23}
gte-Qwen2-1.5B & 2.0 & 4.0 & 3.5 & 4.1 & 3.3 & 3.1 & 1.8 & 2.1 & 0.0 & 0.9 & 0.0 & 5.9 & 9.1 & 5.8 & 6.5 & 4.1 & 8.2 & 7.8 & 8.7 & 3.2 & 3.7 & 0.0 \\
jina-code-1.5b & 18.3 & 23.1 & 14.1 & 18.2 & 15.9 & 14.6 & 16.2 & 10.5 & 12.2 & 9.6 & 0.0 & 33.1 & 46.0 & 23.8 & 37.1 & 31.0 & 31.4 & 35.5 & 27.2 & 21.1 & 19.6 & 0.0 \\
\cmidrule(lr){1-23}
\rowcolor{black!4}\multicolumn{23}{c}{\textbf{\textit{Large Embedding Models \hspace{0.35em}($\geq$4B)}}}\\
\cmidrule(lr){1-23}
Qwen3-4B & 19.3 & 15.1 & 16.0 & 15.6 & 19.3 & 10.0 & 18.4 & 12.5 & 4.2 & 4.4 & 0.0 & 38.8 & 33.4 & 25.4 & 33.5 & 32.2 & 28.0 & 37.8 & 30.7 & 17.9 & 11.3 & 0.0 \\
C2LLM-7B & 16.6 & 24.1 & 20.1 & 19.4 & 19.1 & 12.1 & 17.3 & 10.3 & 11.3 & 2.2 & 38.9 & 33.7 & 46.1 & 32.5 & 40.4 & 30.8 & 35.9 & 40.8 & 28.1 & 27.3 & 14.6 & 23.6 \\
e5-mistral-7b & 20.1 & 25.5 & 15.2 & 16.3 & 19.0 & 11.3 & 16.2 & 14.5 & 9.6 & 6.6 & 0.0 & 38.7 & 45.6 & 20.6 & 31.3 & 29.1 & 27.2 & 34.3 & 31.6 & 33.2 & 18.2 & 0.0 \\
SweRankEmbed-L & 25.9 & 26.8 & 21.2 & 18.0 & 25.0 & 10.6 & 17.9 & 8.7 & 9.2 & \snd{18.7} & 0.0 & 42.6 & 44.3 & 24.0 & 35.0 & 34.9 & 25.9 & 33.9 & 23.5 & 23.2 & 19.4 & 0.0 \\
F2LLM-v2-8B & 20.5 & 17.1 & 14.7 & 19.0 & 17.6 & 13.2 & 16.4 & 13.4 & 10.6 & 10.6 & 38.7 & 36.9 & 38.2 & 21.3 & 34.3 & 32.3 & 35.8 & 36.2 & 33.0 & 29.1 & 24.2 & 14.8 \\
Qwen3-8B & 21.9 & 14.6 & 18.6 & 22.1 & 22.6 & 12.5 & 21.0 & 17.0 & 13.4 & 4.2 & 45.6 & 39.4 & 32.2 & 23.5 & 40.6 & 33.4 & 32.1 & 45.6 & 35.9 & 31.9 & 13.9 & \snd{32.3} \\
\cmidrule(lr){1-23}
\rowcolor{black!4}\multicolumn{23}{c}{\textbf{\textit{In-domain SFT Models \hspace{0.35em}(0.6B--8B)}}}\\
\cmidrule(lr){1-23}
Qwen3-0.6B-SFT & 27.5 & 33.3 & 31.8 & 28.1 & 32.3 & 17.6 & 25.4 & 19.0 & 12.2 & 15.5 & 48.2 & 48.8 & 56.7 & 43.3 & 46.8 & 46.4 & 44.7 & 48.6 & 40.4 & 38.7 & 25.3 & 18.5 \\
Qwen3-4B-SFT & \snd{30.6} & \snd{36.3} & \snd{36.6} & \snd{33.4} & \snd{36.0} & \snd{19.7} & \snd{26.5} & \snd{22.7} & \snd{26.5} & 14.9 & \bst{92.0} & \snd{51.6} & \snd{61.6} & \snd{50.7} & \snd{50.4} & \bst{53.4} & \snd{45.7} & \snd{50.3} & \snd{47.1} & \snd{50.3} & \snd{30.2} & \bst{38.3} \\
Qwen3-8B-SFT & \bst{33.7} & \bst{38.8} & \bst{39.8} & \bst{33.6} & \bst{36.8} & \bst{21.8} & \bst{29.1} & \bst{27.8} & \bst{27.7} & \bst{19.4} & \snd{60.5} & \bst{53.7} & \bst{64.3} & \bst{51.9} & \bst{52.2} & \snd{51.5} & \bst{48.1} & \bst{52.6} & \bst{48.3} & \bst{53.1} & \bst{30.4} & 23.2 \\
\bottomrule
\end{tabular*}
\end{table*}

\begin{table}[!t]
\centering
\scriptsize
\caption{%
  \textbf{\benchname{} LEVEL-2 results for a subset of models by repository difficulty (NDCG@10/Recall@100).}
  \textbf{F/D} denote focused/dense relevance regimes, and \textbf{S/L} denote small/large corpus regimes.
  Full LEVEL-2/3 difficulty results are provided in Table~\ref{tab:difficulty-results-full} in Appendix~\ref{app:detailed-results}.
}
\label{tab:difficulty-results}
\setlength{\tabcolsep}{2pt}
\renewcommand{\arraystretch}{1.05}
\begin{tabular*}{\columnwidth}{@{\extracolsep{\fill}}lcccc@{}}
\toprule
\textbf{Model} & \textbf{F-S} & \textbf{F-L} & \textbf{D-S} & \textbf{D-L} \\
\cmidrule(lr){1-5}
\rowcolor{black!4}\multicolumn{5}{c}{\textbf{\textit{Small Embedding Models \hspace{0.35em}(<1B)}}}\\
\cmidrule(lr){1-5}
CodeRankEmbed & 13.4/43.7 & 9.5/28.8 & 14.5/39.7 & 12.6/28.4 \\
bge-m3 & 4.7/25.8 & 3.5/13.5 & 6.9/26.2 & 5.3/12.9 \\
Qwen3-0.6B & 17.4/54.3 & 16.1/45.6 & 18.6/46.6 & 17.8/40.9 \\
\cmidrule(lr){1-5}
\rowcolor{black!4}\multicolumn{5}{c}{\textbf{\textit{Medium Embedding Models \hspace{0.35em}(1B--2B)}}}\\
\cmidrule(lr){1-5}
gte-Qwen2-1.5B & 2.9/21.5 & 1.7/8.6 & 6.5/25.1 & 3.4/12.1 \\
jina-code-1.5b & 16.0/58.0 & 17.6/49.8 & 19.2/48.9 & 18.9/43.0 \\
\cmidrule(lr){1-5}
\rowcolor{black!4}\multicolumn{5}{c}{\textbf{\textit{Large Embedding Models \hspace{0.35em}($\geq$4B)}}}\\
\cmidrule(lr){1-5}
Qwen3-4B & 16.9/55.2 & 16.0/45.3 & 19.9/48.8 & 18.6/41.2 \\
C2LLM-7B & 18.4/53.0 & 16.6/46.0 & 20.6/47.4 & 19.4/41.0 \\
e5-mistral-7b & 19.8/60.6 & 18.4/51.3 & 21.4/52.1 & 19.9/44.1 \\
SweRankEmbed-L & 24.1/61.7 & 23.1/50.8 & 22.5/51.7 & 24.0/43.8 \\
F2LLM-v2-8B & 14.2/49.1 & 17.9/44.5 & 21.3/47.3 & 20.2/37.9 \\
Qwen3-8B & 18.8/55.9 & 19.0/48.2 & 21.8/48.3 & 20.6/41.7 \\
\cmidrule(lr){1-5}
\rowcolor{black!4}\multicolumn{5}{c}{\textbf{\textit{In-domain SFT Models \hspace{0.35em}(0.6B--8B)}}}\\
\cmidrule(lr){1-5}
Qwen3-0.6B-SFT & 28.2/67.9 & 28.0/62.1 & 26.0/55.4 & 29.2/54.2 \\
Qwen3-4B-SFT & \cellcolor{second}\underline{30.9}/\textbf{74.7} & \cellcolor{second}\underline{31.8}/\underline{69.2} & \cellcolor{second}\underline{29.1}/\textbf{61.2} & \cellcolor{second}\underline{32.3}/\underline{59.4} \\
Qwen3-8B-SFT & \cellcolor{best}\textbf{34.0}/\underline{73.2} & \cellcolor{best}\textbf{35.0}/\textbf{69.7} & \cellcolor{best}\textbf{32.7}/\underline{60.5} & \cellcolor{best}\textbf{33.7}/\textbf{60.4} \\
\bottomrule
\end{tabular*}
\end{table}

\subsection{In-domain SFT Performance}
The last group of Table~\ref{tab:overall} reports the results of embedding models trained with in-domain supervision. We find that simple SFT substantially improves retrieval performance on LEVEL-2/3. Qwen3-0.6B-SFT already outperforms larger embedding models on both LEVEL-2 and LEVEL-3, and Qwen3-8B-SFT achieves the best overall performance on code retrieval in agentic coding settings. Nevertheless, there remains a clear gap between its LEVEL-2/3 performance and its LEVEL-1 performance. This indicates that supervision from pull requests helps the model learn repository-level relevance signals, but it still does not fully solve the difficulty of multi-target change localization.

\subsection{Cross-level Correlation Analysis}
Figure~\ref{fig:level-correlation} gives the cross-level correlation analysis. The LEVEL-1 average only moderately predicts LEVEL-2 performance. Code-to-text retrieval is a relatively weak proxy for both agentic levels, while the hybrid task has the strongest correlation. Together with the absolute performance drop in Table~\ref{tab:overall}, this shows that LEVEL-2 and LEVEL-3 preserve part of the model-ranking signal from traditional retrieval tasks, but also introduce additional challenges related to repository-local structure and requirement-conditioned context.

\subsection{Performance on Different Languages}
Table~\ref{tab:language-results} reports results by programming language. The benefit of in-domain SFT is not limited to Python. On LEVEL-2, Qwen3-8B-SFT obtains the best NDCG@10 on most languages and substantially improves over the non-adapted Qwen3-8B baseline. The improvement is especially clear on JavaScript, TypeScript, and Go. This suggests that repository-local naming conventions, framework structure, and engineering patterns matter for requirement-conditioned retrieval beyond generic semantic similarity.
Language difficulty cannot be explained only by popularity. Even after SFT, Java remains one of the harder LEVEL-2 tasks. Overall, task-specific adaptation improves cross-language robustness, but repository-local requirement-conditioned retrieval remains uneven across ecosystems.

\subsection{Performance across Difficulty Levels}
Table~\ref{tab:difficulty-results} reports LEVEL-2 performance under different repository difficulties. We draw two conclusions: (1) In-domain SFT improves performance in all difficulty settings, including the hardest dense-large setting. (2) Even for the strongest SFT model, Recall@100 still drops when moving from focused or smaller settings to dense repositories. This indicates that when the candidate pool becomes larger and the repository contains many similar distractors, relevant edit chunks can still be pushed out of the retrieved set. 
Besides, SFT models maintain relatively stable NDCG@10 across difficulty settings, suggesting that the adapted models learn useful repository-local ranking signals once relevant candidates are retrieved. Therefore, dense-large repositories form an important stress test for future retrievers.

\begin{table}[t]
\centering
\footnotesize
\caption{LEVEL-2/3 target expansion examples.}
\label{tab:main-target-expansion-cases}
\setlength{\tabcolsep}{3pt}
\renewcommand{\arraystretch}{1.1}
\begin{tabularx}{\columnwidth}{@{}L{0.30\columnwidth}Y@{}}
\toprule
\textbf{Instance} & \textbf{Signal and Targets} \\
\midrule
\texttt{django-15499} & \textbf{Request:} optimize \texttt{CreateModel + AlterModelManagers} into \texttt{CreateModel}. \newline \textbf{L2:} 1 qrel for the migration operation-reduction branch. \newline \textbf{L3:} 58 qrels covering the migration graph, duplicate-object validation, model-state objects, and helper operations. \\
\midrule
\texttt{grpc-go-3201} & \textbf{Request:} do not call \texttt{NewServiceConfig} when DNS lookups are disabled. \newline \textbf{L2:} 2 qrels for DNS resolver and resolver-wrapper callers. \newline \textbf{L3:} 121 qrels over the resolver stack, \texttt{ClientConn} state updates, default service-config behavior, and wrapper logic. \\
\bottomrule
\end{tabularx}
\end{table}

\begin{table}[t]
\centering
\footnotesize
\caption{Query rewrite examples.}
\label{tab:main-query-rewrite-cases}
\setlength{\tabcolsep}{3pt}
\renewcommand{\arraystretch}{1.1}
\begin{tabularx}{\columnwidth}{@{}L{0.30\columnwidth}Y@{}}
\toprule
\textbf{Instance} & \textbf{Rewrite Effect} \\
\midrule
\texttt{streamlink-\allowbreak5926} & \textbf{Raw:} 1,904-character bug report with checklist text, Kuwaiti channel URLs, and debug logs. \newline \textbf{Rewrite:} keeps the ``No plugin can handle URL'' failure and concrete media.gov.\allowbreak kw URLs as retrieval anchors. \\
\midrule
\texttt{fmt-3729} & \textbf{Raw:} verbose discussion of generic versus native formatting for filesystem paths. \newline \textbf{Rewrite:} requests generic and native format specifiers while preserving slash-direction behavior. \\
\midrule
\texttt{camunda-27226} & \textbf{Raw:} change the \texttt{state} field from string to enum in ProcessInstance and ProcessInstance\allowbreak Filter. \newline \textbf{Rewrite:} keeps class names, \texttt{state}, and OpenAPI enum values while turning the note into a developer question. \\
\bottomrule
\end{tabularx}
\end{table}

\subsection{Qualitative Case Studies}
Tables~\ref{tab:main-target-expansion-cases} and~\ref{tab:main-query-rewrite-cases} highlight representative cases of \benchname{}. LEVEL-3 needs broader recall because useful context can be spread across state-management, validation, and helper paths rather than concentrated at the final edit site. Query rewriting removes boilerplate and changes the lexical and structural signals available to retrievers. These examples clarify why the two agentic levels reward different retrieval behavior. 
In \texttt{django-15499}, the direct edit target is narrow, but the agent also needs migration-graph and model-state context to judge whether the optimization is valid. In \texttt{grpc-go-3201}, the useful context expands from resolver call sites to \texttt{ClientConn} state and service-configuration propagation. Thus, LEVEL-3 is a context-coverage setting rather than a looser version of edit localization. The rewrite cases further show that cleaner queries are useful only when they preserve repository anchors, such as concrete URLs, error strings, API names, fields, and enum values; otherwise, rewriting can remove exactly the lexical cues that help retrieve local code amid cross-file distractors during realistic agent runs. 
Full cases are provided in Tables~\ref{tab:case-level2-level3} and~\ref{tab:case-rewrite-contrast} in Appendix~\ref{app:data-case-studies}.

\section{Conclusion}

In this paper, we presented \textbf{\benchname{}}, a comprehensive benchmark for code retrieval in agentic coding. By organizing evaluation from traditional code understanding to issue-to-edit localization and broader context retrieval, \benchname{} exposes the gap between snippet-level code search and requirement-driven repository search. Our experiments show that current general embedding models and code-oriented embedding models degrade sharply in this setting. Besides, in-domain SFT with pull request data across different open-source repositories substantially improves LEVEL-2/3 retrieval of \benchname{}.

\section*{Limitations}

\benchname{} is mainly built from SWE-bench-series and SWE-bench-style software-change data. It may therefore not fully capture retrieval on newly created or rapidly evolving repositories, and it does not yet include freshness-oriented metrics for such settings. Its diversity is also uneven: popular ecosystems contribute more examples, while long-tail languages and uncommon project structures remain underrepresented.
\bibliography{custom_checked}

\appendix

\section{Detailed Dataset Statistics}
\label{app:dataset-detail-stats}

Table~\ref{tab:dataset_stats_detail} reports the sub-dataset-level statistics
for all three \benchname{} levels behind the aggregate summary in
Table~\ref{tab:dataset_stats}.

\paragraph{LEVEL-1.}
LEVEL-1 brings together standard code-understanding retrieval tasks rather than
repository-local requirement data. The rows differ noticeably in query style and
relevance density: CodeSearchNet-style tasks dominate the corpus scale, CLARC
uses much denser relevance labels, and CosQA or Text2SQL-style tasks contain
shorter natural-language queries. This level is mainly a reference point for
general code-semantic matching.

\paragraph{LEVEL-2.}
LEVEL-2 moves to issue-to-edit localization inside concrete repository
snapshots. The corpus is much larger, while each request is linked to a relatively
small set of edit-bearing chunks. Different SWE-bench-derived sources vary in
repository coverage and target density, but the common difficulty is the same:
the retriever must find the code that is likely to be edited, not just broadly
related code.

\paragraph{LEVEL-3.}
LEVEL-3 keeps the requirement-centered setting but expands the target from edit
locations to useful surrounding context. Its relevance sets are therefore
denser than LEVEL-2, especially for cases where completing the change requires
nearby helpers, tests, state definitions, or related implementation paths. This
level is intended to evaluate whether retrieval can support an agent's broader
context gathering, not only the first edit location.

\benchname{} is constructed from publicly released code-retrieval datasets,
SWE-bench-series benchmarks, and open-source GitHub repositories. We use only
data sources that are available for research or benchmark use, follow the
corresponding upstream licenses and terms, and do not include private or
proprietary repositories. This construction avoids introducing additional
license risk beyond the original open-source data sources.

\begin{table*}[t]
\centering
\caption{Detailed \benchname{} dataset statistics. Rewrite variants share the
same corpus and qrels with their non-rewrite counterparts. Total-row averages
are weighted by the corresponding total query or corpus counts.}
\label{tab:dataset_stats_detail}
\scriptsize
\setlength{\tabcolsep}{1.8pt}
\renewcommand{\arraystretch}{1.06}
\resizebox{\textwidth}{!}{%
\begin{tabular}{@{}l*{8}{c}@{}}
\toprule
\textbf{Sub-dataset}
 & \textbf{\#Repos}
 & \textbf{\#Queries}
 & \textbf{\#Corpus}
 & \textbf{\#Qrels}
 & \textbf{Rels/Q}
 & \textbf{Avg.Q Len}
 & \textbf{Avg.Rw.Q Len}
 & \textbf{Avg.C Len}
\\
\cmidrule(lr){1-9}
\rowcolor{black!4}\multicolumn{9}{c}{\textbf{\textit{LEVEL-1: Code Understanding}}}\\
\cmidrule(lr){1-9}
CPRet & -- & 19{,}806 & 102{,}259 & 91{,}558 & 4.6 & 1{,}464 & -- & 2{,}641 \\
CLARC & -- & 6{,}210 & 7{,}228 & 584{,}842 & 94.2 & 431 & -- & 971 \\
AppsRetrieval & -- & 3{,}765 & 8{,}765 & 3{,}765 & 1.0 & 1{,}670 & -- & 575 \\
CodeFeedbackST & -- & 31{,}306 & 156{,}526 & 31{,}306 & 1.0 & 724 & -- & 1{,}521 \\
CodeSearchNetCCRetrieval & -- & 52{,}561 & 1{,}005{,}474 & 52{,}561 & 1.0 & 388 & -- & 267 \\
COIRCodeSearchNetRetrieval & -- & 52{,}561 & 1{,}003{,}765 & 52{,}561 & 1.0 & 663 & -- & 183 \\
CodeTransOceanContest & -- & 221 & 1{,}008 & 221 & 1.0 & 1{,}012 & -- & 1{,}508 \\
CodeTransOceanDL & -- & 180 & 816 & 180 & 1.0 & 1{,}868 & -- & 1{,}479 \\
CosQA & -- & 500 & 20{,}604 & 500 & 1.0 & 37 & -- & 276 \\
SyntheticText2SQL & -- & 5{,}851 & 105{,}851 & 5{,}851 & 1.0 & 83 & -- & 127 \\
\textbf{LEVEL-1 Total} & \textbf{--} & \textbf{172{,}961} & \textbf{2{,}412{,}296} & \textbf{823{,}345} & \textbf{4.8} & \textbf{676} & \textbf{--} & \textbf{412} \\
\cmidrule(lr){1-9}
\rowcolor{black!4}\multicolumn{9}{c}{\textbf{\textit{LEVEL-2: Issue-to-Edit Localization}}}\\
\cmidrule(lr){1-9}
SWE-bench Verified      & 12  & 432    & 398{,}397   & 1{,}173  & 2.72  & 1{,}685.9 & 772.5 & 1{,}339.8 \\
SWE-bench Pro           & 10  & 632    & 586{,}719   & 8{,}088  & 12.80 & 1{,}292.9 & 528.2 & 1{,}316.1 \\
SWE-bench Live          & 209 & 1{,}623 & 1{,}229{,}695 & 14{,}028 & 8.64  & 2{,}237.3 & 811.1 & 1{,}388.3 \\
SWE-bench++             & 319 & 442    & 5{,}365{,}829 & 6{,}197  & 14.02 & 1{,}564.8 & 756.1 & 830.5 \\
SWE-bench+              & 6   & 207    & 57{,}320    & 1{,}483  & 7.16  & 2{,}027.4 & 849.8 & 1{,}576.9 \\
Multi-SWE-bench         & 35  & 1{,}449 & 858{,}042   & 20{,}384 & 14.07 & 724.6 & 418.2 & 1{,}406.2 \\
SWE-bench Multilingual  & 41  & 276    & 881{,}118   & 1{,}359  & 4.92  & 1{,}991.5 & 884.9 & 745.1 \\
\textbf{LEVEL-2 Total}  & \textbf{632} & \textbf{5{,}061} & \textbf{9{,}377{,}120} & \textbf{52{,}712} & \textbf{10.42} & \textbf{1{,}558.5} & \textbf{660.8} & \textbf{1{,}004.9} \\
\cmidrule(lr){1-9}
\rowcolor{black!4}\multicolumn{9}{c}{\textbf{\textit{LEVEL-3: Broader Context Retrieval}}}\\
\cmidrule(lr){1-9}
SWE-bench Verified      & 12  & 302    & 372{,}258   & 8{,}928  & 29.56 & 1{,}562.7 & 799.1 & 1{,}319.4 \\
SWE-bench Pro           & 10  & 585    & 509{,}200   & 14{,}987 & 25.62 & 1{,}292.1 & 614.1 & 1{,}319.8 \\
Multi-SWE-bench         & 35  & 1{,}445 & 857{,}861   & 77{,}293 & 53.49 & 724.3 & 416.7 & 1{,}406.2 \\
SWE-bench Multilingual  & 40  & 248    & 870{,}262   & 5{,}271  & 21.25 & 2{,}006.4 & 845.6 & 740.1 \\
\textbf{LEVEL-3 Total}  & \textbf{97} & \textbf{2{,}580} & \textbf{2{,}609{,}581} & \textbf{106{,}479} & \textbf{41.27} & \textbf{1{,}074.4} & \textbf{547.4} & \textbf{1{,}154.8} \\
\bottomrule
\end{tabular}}
\end{table*}

\section{SFT Training Details}
\label{app:sft-training-details}
\label{app:training-data-stats}

Table~\ref{tab:training-data-scale} reports the scale of the in-domain training
data used for continued embedding-model fine-tuning. The training set covers
628 repositories and 53{,}301 queries collected from pull requests, yielding 1{,}218{,}875
relevance labels over a corpus of 29{,}171{,}451 chunks. On average, each query
has 22.87 relevant chunks, indicating that the supervision is not limited to
single-location matching but often contains multiple patch-aligned code
targets. The average query length is 375.1 characters, while the average corpus
chunk length is 2{,}188.4 characters, giving the model relatively concise PR
descriptions paired with substantially longer repository-local code contexts.

\begin{table}[t]
\centering
\caption{Training data scale.}
\label{tab:training-data-scale}
\setlength{\tabcolsep}{4.2pt}
\small
\renewcommand{\arraystretch}{1.15}
\begin{tabular*}{\columnwidth}{@{\extracolsep{\fill}}lc@{}}
\toprule
\textbf{Metric} & \textbf{Value} \\
\midrule
\# Repos & 628 \\
\# Queries & 53{,}301 \\
\# Qrels & 1{,}218{,}875 \\
\# Corpus & 29{,}171{,}451 \\
Avg. query length & 375.1 \\
Avg. corpus length & 2{,}188.4 \\
Qrels / Query & 22.87 \\
\bottomrule
\end{tabular*}
\end{table}

We fine-tune the Qwen3-Embedding family with in-domain supervision derived
from LEVEL-2-style PR data. The repositories are disjoint from the SWE-bench
family. For each PR, we use the PR description as the query and patch-aligned
chunks as positives. For each query, 64 non-positive chunks are sampled from
the same repository to form a repository-local negative pool. During training,
examples are drawn with a 1:8 positive-to-negative ratio, so the model must
separate true edit targets from plausible distractors in the same codebase.
All in-domain SFT rows in the experiments report checkpoints after three
supervised fine-tuning epochs.

For each query $q$ with positive chunks $\mathcal{P}$ and repository-local
negatives $\mathcal{N}$, we optimize a multi-positive InfoNCE loss:
\begin{equation}
\mathcal{L}(q) =
-\log
\frac{\sum_{d \in \mathcal{P}} \exp(s_\theta(q,d)/\tau)}
{\sum_{d \in \mathcal{P}\cup\mathcal{N}} \exp(s_\theta(q,d)/\tau)}.
\label{eq:sft-infonce}
\end{equation}
where $s_\theta(q,d)$ is the query--document embedding similarity and $\tau$
is the temperature.

\section{Details of All Baselines}
\label{app:baseline-models}

Table~\ref{tab:overall} intentionally includes both general embedding
models and code-oriented retrieval models. General-purpose models are used to
test whether broad semantic representations transfer to agentic code retrieval,
whereas code-specific models are used to test whether supervision from code
search, code understanding, or issue localization better matches the task
structure of \benchname{}. When a table row is a fine-tuned variant, the citation
points to the public base model family; the fine-tuning itself is the in-domain
adaptation described in Appendix~\ref{app:sft-training-details}. All
in-domain SFT rows use checkpoints after three supervised fine-tuning epochs.

\paragraph{Lexical baseline.}
\begin{itemize}[noitemsep,topsep=2pt]
\item \textbf{BM25}~\cite{BM25} is a sparse lexical retrieval baseline based
on exact term matching, inverse document frequency, and document-length
normalization. 
\end{itemize}

\paragraph{General embedding models.}
\begin{itemize}[noitemsep,topsep=2pt]
\item \textbf{multilingual-e5-small/base/large}~\cite{multilingual-e5} are
encoder-only multilingual-E5 checkpoints trained as general text embedding
models over multilingual contrastive data.
\item \textbf{bge-large-en-v1.5}~\cite{bge-large-en-v1.5} is a model from the
BGE embedding family.
\item \textbf{bge-m3}~\cite{bge-m3} extends BGE toward multilingual,
multifunctional, and multi-granularity retrieval. It is still a general
embedding model, but its long-input and hybrid-retrieval design makes it
relevant for large code-context evaluation.
\item \textbf{gte-multilingual-base}~\cite{gte-multilingual-base} is an
encoder-only multilingual GTE model with long-context support and dense/sparse
representation capability.
\item \textbf{jina-emb-v5-small}~\cite{jina-embeddings-v5-text-small} refers
to \texttt{jina-embeddings-v5-text-small}, a compact task-targeted text
embedding model distilled for efficient general-purpose retrieval.
\item \textbf{gte-Qwen2-1.5B-instruct}~\cite{gte-Qwen2-1.5B-instruct} is an
LLM-based GTE model built on Qwen2-1.5B and instruction-tuned for retrieval.
\item \textbf{F2LLM-v2-1.7B/4B/8B}~\cite{F2LLM-v2} are general-purpose
multilingual embedding LLMs from the F2LLM-v2 family.
\item \textbf{e5-mistral-7b-instruct}~\cite{e5-mistral-7b-instruct} fine-tunes
a Mistral-7B backbone for text embedding with synthetic and labeled retrieval
data. It is a strong general-purpose LLM-based embedding baseline.
\item \textbf{pplx-embed-v1-4b}~\cite{pplx-embed-v1-4b} is a Perplexity
embedding model trained for web-scale multilingual retrieval on a
diffusion-pretrained backbone.
\item \textbf{Qwen3-Embedding-0.6B/4B/8B}~\cite{Qwen3-Embedding} are public
Qwen3-Embedding models. They provide a controlled general-embedding family for
testing how model scale affects zero-shot transfer to agentic code retrieval
before any in-domain adaptation.
\end{itemize}

\paragraph{Code-oriented models.}
\begin{itemize}[noitemsep,topsep=2pt]
\item \textbf{CodeRankEmbed}~\cite{CodeRankEmbed} is a 137M bi-encoder trained
for code retrieval using contrastive data from CoRNStack. It is a compact
code-specialized baseline for aligning natural-language queries with code.
\item \textbf{jina-code-emb-0.5b/1.5b}~\cite{jina-code-embeddings} are Jina
Code Embeddings checkpoints built from Qwen2.5-Coder backbones. They support
multiple code retrieval directions, including natural-language-to-code and
code-to-code.
\item \textbf{C2LLM-7B}~\cite{C2LLM-7B} is a code contrastive LLM for code
retrieval with adaptive cross-attention pooling.
\item \textbf{SweRankEmbed-Large}~\cite{SweRankEmbed-Large} is a 7B
issue-localization retriever trained to map GitHub issues and code snippets
into a shared retrieval space.
\item \textbf{Qwen3-Embedding-0.6B/4B/8B-ft}~\cite{Qwen3-Embedding} are our
in-domain fine-tuned Qwen3 Embedding variants. They keep the base model family
fixed while adapting different model scales to \benchname{}-style request-to-code
retrieval data under the same benchmark protocol.
\end{itemize}

\section{Evaluation Metrics}
\label{app:evaluation-metrics}

\benchname{} reports NDCG@10 and Recall@100 for retrieval evaluation. Unless
otherwise specified, result tables report each cell as NDCG@10/Recall@100.

\paragraph{NDCG@10.}
NDCG@10 measures whether highly relevant chunks appear near the top of the
retrieved list. For a query \(q\), let \(rel_i\) be the relevance score of the
chunk ranked at position \(i\). We compute
\[
\mathrm{DCG@10}(q)=
\sum_{i=1}^{10}\frac{2^{rel_i}-1}{\log_2(i+1)}.
\]
\[
\mathrm{NDCG@10}(q)=
\frac{\mathrm{DCG@10}(q)}{\mathrm{IDCG@10}(q)}.
\]
This metric is especially important for LEVEL-2, where the retriever should
place edit-bearing chunks among the first results an agent is likely to inspect.

\paragraph{Recall@100.}
Recall@100 measures coverage of the relevant context. Let \(R_q\) be the set of
positive-relevance chunks for query \(q\), and let \(T_{100}(q)\) be the top-100
retrieved chunks. We compute
\[
\mathrm{Recall@100}(q)=\frac{|R_q \cap T_{100}(q)|}{|R_q|}.
\]
This metric is especially important for LEVEL-3, where useful information may
be spread across edit locations, helper code, tests, configuration, and
documentation.

\section{LEVEL-3 Annotation Pipeline}
\label{app:level3-annotation-pipeline}

Algorithm~\ref{alg:level3-annotation} summarizes the automated annotation
procedure used to construct LEVEL-3. The pipeline is designed to recover
context that helps an agent understand and localize a requested change, rather than only
the chunks eventually edited by a patch. The allowlist rerun is therefore used
as a functional validation signal, while final qrels are still determined by
trajectory extraction and relevance judging.

\paragraph{Notation.}
Here, \(\mathcal{I}\) denotes the issue set, \(\mathcal{C}_i\) is the chunked
repository corpus for issue \(i\), and \(\mathcal{Q}^{(2)}_i\) and
\(\mathcal{Q}^{(3)}_i\) are the LEVEL-2 edit-location qrels and LEVEL-3
broader-context qrels, respectively. The agent is \(\mathcal{A}\), the LLM
judges are \(\mathcal{M}\), and \(\mathcal{U}\) is the set of browse commands
parsed from trajectories. During annotation, \(\tau_i\) is an agent trajectory,
\(\mathcal{B}_i\) is the browsed text extracted from it, \(\mathcal{S}_i\) is
the aligned candidate chunk set, and \(\mathcal{R}_i\) contains chunks accepted
by relevance judging. The allowlists \(\mathcal{W}^{(2)}_i\) and
\(\mathcal{W}^{(3)}_i\) are built from LEVEL-2 and LEVEL-3 qrels, and
\((v^{(2)}_i,v^{(3)}_i)\) record the corresponding validation outcomes.

\begin{algorithm}[H]
\caption{LEVEL-3 annotation pipeline.}
\label{alg:level3-annotation}
\footnotesize
\begin{tabularx}{\linewidth}{@{}>{\bfseries}l@{\hspace{0.45em}}Y@{}}
Input: &
$\mathcal{I}$, $\{\mathcal{C}_i\}$, $\{\mathcal{Q}^{(2)}_i\}$,
agent $\mathcal{A}$, judges $\mathcal{M}$, commands
\(\mathcal{U}\) = \{\texttt{cat}, \texttt{grep}, \texttt{head},
\texttt{sed}\}, budget $B$. \\
Output: &
$\{\mathcal{Q}^{(3)}_i\}$ and validation rates
$\{(v^{(2)}_i,v^{(3)}_i)\}$. \\
\end{tabularx}

\vspace{2pt}
\setlength{\tabcolsep}{0pt}
\renewcommand{\arraystretch}{1.03}
\begin{tabularx}{\linewidth}{@{}r@{\hspace{0.65em}}Y@{}}
1  & \textbf{for} each issue $i\in\mathcal{I}$ \textbf{do} \\
2  & \quad $\mathcal{Q}^{(3)}_i \leftarrow \mathcal{Q}^{(2)}_i$;
      $\mathcal{S}_i \leftarrow \varnothing$; $t\leftarrow 0$ \\
3  & \quad \textbf{repeat} \\
4  & \qquad $(\hat{p}_i,r_i,\tau_i) \leftarrow$
      \textsc{RunAgent}$(\mathcal{A}, i)$ \\
5  & \qquad $\mathcal{B}_i \leftarrow$
      \textsc{ExtractBrowse}$(\tau_i,\mathcal{U})$ \\
6  & \qquad $\mathcal{S}_i \leftarrow \mathcal{S}_i \cup
      {}$ \textsc{AlignToChunks}$(\mathcal{B}_i,\mathcal{C}_i)$ \\
7  & \qquad $\mathcal{R}_i \leftarrow$
      \textsc{VoteJudge}$(\mathcal{M},q_i,\mathcal{S}_i)$ \\
8  & \qquad $\mathcal{Q}^{(3)}_i \leftarrow
      \mathcal{Q}^{(3)}_i \cup
      \{c \in \mathcal{R}_i \mid \mathrm{score}(c)>0\}$ \\
9  & \qquad $t \leftarrow t+1$ \\
10 & \quad \textbf{until} $r_i=1$ \textbf{or}
      \textsc{NoNewContext}$(\mathcal{Q}^{(3)}_i)$ \textbf{or} $t=B$ \\
11 & \quad $\mathcal{W}^{(2)}_i \leftarrow$
      \textsc{BuildAllowlist}$(\mathcal{Q}^{(2)}_i)$;
      \\
12 & \quad $\mathcal{W}^{(3)}_i \leftarrow$
      \textsc{BuildAllowlist}$(\mathcal{Q}^{(3)}_i)$ \\
13 & \quad $(v^{(2)}_i,v^{(3)}_i) \leftarrow$
      \textsc{AllowlistEval}$(\mathcal{A},\mathcal{W}^{(2)}_i,
      \mathcal{W}^{(3)}_i)$ \\
14 & \textbf{end for} \\
15 & \textbf{return} $\{\mathcal{Q}^{(3)}_i\}_{i\in\mathcal{I}}$ and
      aggregate validation statistics over $\{(v^{(2)}_i,v^{(3)}_i)\}$.
\end{tabularx}

\vspace{2pt}
\end{algorithm}

\section{Case Studies}
\label{app:data-case-studies}

The following cases illustrate how \benchname{} cleans raw SWE-bench-style
instances and reshapes them into agentic-code-search queries. We use compact
tables to show the source signal, filtering decision, query type, and rewrite
effect side by side.

\subsection{Level-2 and Level-3 Target Expansion}
\label{app:case-level2-level3}

For overlapping instances, LEVEL-2 focuses on edit-bearing chunks, whereas LEVEL-3 expands the target
into surrounding implementation, data-flow, and test context that can help a
coding agent reason about the fix. Table~\ref{tab:case-level2-level3} shows
that this expansion is not a small formatting change: one or two direct edit
chunks can become dozens of context chunks once tests, helper paths, resolver
state, or migration state need to be retrieved. This is why LEVEL-3 is treated
as a broader context-retrieval task rather than only a relaxed version of
LEVEL-2.

\subsection{Level-3 Annotation Quality}
\label{app:case-level3-quality}

The allowlist evaluation in Section~\ref{sec:level3-construction} checks
whether the added LEVEL-3 context is useful for completing requested changes, rather than
only textually related to the query. Table~\ref{tab:level3-allowlist-quality}
reports the resolve rate under three settings. The Full Agent column is the
original mini-swe-agent run used during annotation, with normal repository
access. The LEVEL-2 and LEVEL-3 columns rerun the agent in our allowlist
environment, where readable files are restricted to the corresponding qrels.
Across all sources, the LEVEL-3 allowlist consistently solves more instances than
the LEVEL-2 allowlist, showing that the additional context labels recover
information needed by the agent beyond the final edit locations.

These rates should be read as a relative validation signal. Our allowlist
environment restricts file reading and file operations more strongly than the
standard mini-swe-agent setting, and these restrictions can also affect the
agent's testing step. As a result, the allowlisted resolve rates can be lower
than the original full-agent resolve rate even when the annotation contains
useful context. The key comparison is therefore between the LEVEL-3 and LEVEL-2
allowlists.

We also do not remove every unresolved trajectory when constructing LEVEL-3.
Keeping only resolved runs would bias the dataset toward simpler tasks with
fewer edits, because those are the cases agents solve more easily. For harder
tasks, the trajectory may still reveal files, tests, helpers, or data-flow
paths that are useful for understanding and localizing the fix. After trajectory
extraction and LLM relevance judging, we retain such useful snippets as qrels
even when the end-to-end agent run does not fully complete the requested change.

\subsection{Filtered Raw Queries}
\label{app:case-filtered-queries}

Some raw SWE-bench-style requests are unsuitable for retrieval evaluation. We
filter queries that leak the answer location or fix, and queries that contain
almost no problem description beyond an issue number or boilerplate.

Table~\ref{tab:filter-query-stats} summarizes the filtering results for the
SWE-bench-derived sources used in \benchname{}. Overall, 489 of 5{,}550 raw
queries are removed. Most filtered queries leak the answer location or patch
recipe, while Multi-SWE-bench contributes most of the meaningless-query cases.
SWE-bench Verified has the highest filtering rate, and SWE-bench Pro has the
lowest. The examples in Table~\ref{tab:case-filtered-queries} show typical
cases behind these counts.

\subsection{Agentic Coding Query Types}
\label{app:case-query-types}

Agentic coding queries are not only bug reports. They also include feature
requests, refactoring requests, documentation fixes, and open-ended questions
where the desired code change must be inferred from user intent.
Table~\ref{tab:case-query-types} gives representative cases. The examples show
that even short requests can carry different localization signals: bug reports
often name failures or inconsistent behavior, feature requests name the desired
API behavior, and documentation or question-style queries require the retriever
to infer which implementation or documentation paths are relevant.

\subsection{Original Query and Rewrite Contrast}
\label{app:case-rewrite-contrast}

The rewrite variant keeps the same corpus and qrels but converts retained issue
or PR text into a natural developer query. Because this split is built after the
query-quality filtering stage, rewriting focuses on removing boilerplate and
redundant context while preserving concrete failures, API names, and desired
changes. Table~\ref{tab:case-rewrite-contrast}
illustrates the intended tradeoff. Rewriting makes the query closer to what a
developer would type into an assistant, but it also compresses context; this is
why the rewritten split is reported separately rather than replacing the
original raw-query setting.

\begin{table*}[t]
\centering
\caption{Level-2 versus Level-3 target expansion cases.}
\label{tab:case-level2-level3}
\scriptsize
\setlength{\tabcolsep}{3pt}
\renewcommand{\arraystretch}{1.12}
\begin{tabularx}{\textwidth}{L{0.18\textwidth}Y Y Y}
\toprule
\textbf{Instance} & \textbf{Raw Query Signal} & \textbf{Level-2 Target} & \textbf{Level-3 Expansion} \\
\midrule
\texttt{django-15499}
\newline SWE-bench Verified
& Optimize \texttt{CreateModel + AlterModelManagers} into \texttt{CreateModel}; the request asks whether this should mirror the existing \texttt{AlterModelOptions} optimization.
& \textbf{1 qrel}. The direct target is the migration operation-reduction branch that checks model-operation types, model names, and proxy options before returning an empty operation list.
& \textbf{58 qrels}. The target set expands to the migration-operation graph, duplicate-object validation, model state objects, and helper operations needed to judge whether the optimization is safe. \\
\midrule
\texttt{zstd-983}
\newline Multi-SWE-bench
& Increase \texttt{windowLog} from a \texttt{CDict} when source size is known, so dictionary compression can fit both the dictionary and source under a smaller limit.
& \textbf{1 qrel}. The direct target is the CDict compression entry point that reads applied compression parameters from the reference context.
& \textbf{79 qrels}. The expanded set covers pledged-source-size handling, compression-level-to-parameter conversion, and CDict internals such as dictionary buffer, content size, and reference context. \\
\midrule
\texttt{grpc-go-3201}
\newline Multi-SWE-bench
& Do not call \texttt{NewServiceConfig} when DNS lookups are disabled; also ignore related resolver-wrapper calls and improve resolver service-config tests.
& \textbf{2 qrels}. The direct targets are DNS resolver and resolver-wrapper callers that parse and apply service-config updates.
& \textbf{121 qrels}. The target set expands to the resolver stack, \texttt{ClientConn} resolver-state updates, default service-config behavior, and wrapper behavior around disabled lookups. \\
\bottomrule
\end{tabularx}
\end{table*}

\begin{table}[t]
\centering
\caption{LEVEL-3 annotation validation.}
\label{tab:level3-allowlist-quality}
\small
\setlength{\tabcolsep}{3.5pt}
\renewcommand{\arraystretch}{1.08}
\begin{tabular*}{\columnwidth}{@{\extracolsep{\fill}}lccc@{}}
\toprule
\textbf{Source} & \textbf{Full} & \textbf{L3} & \textbf{L2} \\
\midrule
SWE-bench Verified      & 73.0\% & 69.0\% & 62.0\% \\
SWE-bench Multilingual  & 65.0\% & 59.0\% & 43.0\% \\
SWE-bench Pro           & 46.1\% & 43.2\% & 36.9\% \\
Multi-SWE-bench         & 42.0\% & 39.5\% & 23.3\% \\
\bottomrule
\end{tabular*}
\end{table}

\begin{table*}[t]
\centering
\caption{Filtered-query statistics for SWE-bench-derived sources.}
\label{tab:filter-query-stats}
\small
\setlength{\tabcolsep}{4pt}
\renewcommand{\arraystretch}{1.08}
\begin{tabular*}{\textwidth}{@{\extracolsep{\fill}}lrrrrrr@{}}
\toprule
\textbf{Source} & \textbf{Raw} & \textbf{Kept} & \textbf{Filtered} & \textbf{Leak} & \textbf{Meaningless} & \textbf{Rate} \\
\midrule
Multi-SWE-bench        & 1{,}615 & 1{,}449 & 166 & 61  & 105 & 10.3\% \\
SWE-bench++            & 475     & 442     & 33  & 32  & 1   & 6.9\% \\
SWE-bench Live         & 1{,}784 & 1{,}623 & 161 & 160 & 1   & 9.0\% \\
SWE-bench+             & 223     & 207     & 16  & 15  & 1   & 7.2\% \\
SWE-bench Multilingual & 300     & 276     & 24  & 23  & 1   & 8.0\% \\
SWE-bench Pro          & 653     & 632     & 21  & 21  & 0   & 3.2\% \\
SWE-bench Verified     & 500     & 432     & 68  & 68  & 0   & 13.6\% \\
\midrule
\textbf{Total}         & \textbf{5{,}550} & \textbf{5{,}061} & \textbf{489} & \textbf{380} & \textbf{109} & \textbf{8.8\%} \\
\bottomrule
\end{tabular*}
\end{table*}

\begin{table*}[t]
\centering
\caption{Raw queries filtered before benchmark construction.}
\label{tab:case-filtered-queries}
\scriptsize
\setlength{\tabcolsep}{3pt}
\renewcommand{\arraystretch}{1.12}
\begin{tabularx}{\textwidth}{L{0.18\textwidth}L{0.18\textwidth}Y Y}
\toprule
\textbf{Query} & \textbf{Verdict} & \textbf{Why Filtered} & \textbf{Raw Signal} \\
\midrule
\texttt{django-11179}
\newline SWE-bench Verified
& Answer leak
& The raw request names the exact module and line range, then states the required update. Retrieval would be nearly solved before the model reads the repository.
& ``\texttt{delete()} on instances of models without dependencies does not clear PKs'' followed by an explicit pointer to the deletion module, lines 276--281, and the exact line to update. \\
\midrule
\texttt{textual-5795}
\newline SWE-bench Live
& Answer leak
& The raw request gives both the root cause and the patch recipe rather than only describing the observed failure.
& Options added after mount are not displayed because the OptionList line-update routine exits early when the scroll region has zero height; the query states the fix is to refresh layout first. \\
\midrule
\texttt{cli-2251}
\newline Multi-SWE-bench
& Meaningless
& The query contains only an issue reference and gives no observable behavior, expected behavior, error message, API name, or implementation clue.
& ``Fix issue \#2216. This fixes \#2216.'' \\
\midrule
\texttt{svelte-13151}
\newline Multi-SWE-bench
& Meaningless
& The only task signal is a short PR title. Most of the body is submission checklist text and branch guidance, which adds noise but not localization signal.
& ``fix: visit expression for \texttt{svelte:component} references'' plus Svelte 5 rewrite boilerplate, close-reference text, and checkbox instructions for PR submission. \\
\bottomrule
\end{tabularx}
\end{table*}

\begin{table*}[t]
\centering
\caption{Agentic coding query type examples.}
\label{tab:case-query-types}
\scriptsize
\setlength{\tabcolsep}{3pt}
\renewcommand{\arraystretch}{1.12}
\begin{tabularx}{\textwidth}{L{0.16\textwidth}L{0.22\textwidth}Y}
\toprule
\textbf{Type} & \textbf{Instance} & \textbf{Query Signal} \\
\midrule
Bug report
& \texttt{gson-1093}
& The Number overload of JsonWriter can write pseudo-numeric values such as \texttt{NaN} and \texttt{Infinity} in lenient mode, but the double overload rejects them, making semi-numeric double output impossible without boxing. \\
\midrule
Feature request
& \texttt{fmt-3729}
& Support both generic and native formatting for filesystem paths, so callers can control slash direction in cross-platform output. \\
\midrule
Refactoring
& \texttt{sqlfluff-3904}
& Standardize the underscored progress-bar option, the only command-line option using underscores instead of dashes; the request asks whether to rename, defer, or accept both forms. \\
\midrule
Question
& \texttt{Pillow-9083}
& Ask whether Pillow can delete only GPS EXIF data rather than all EXIF data; the docs explain deleting from IFD0 but not from GPS IFDs. \\
\midrule
Documentation
& \texttt{pylint-10045}
& The \texttt{--colorized} option for \texttt{pyreverse} is missing from 3.2.0-dev0 documentation even though downstream tools have started integrating it. \\
\bottomrule
\end{tabularx}
\end{table*}

\begin{table*}[t]
\centering
\caption{Original raw queries and rewritten developer queries.}
\label{tab:case-rewrite-contrast}
\scriptsize
\setlength{\tabcolsep}{3pt}
\renewcommand{\arraystretch}{1.12}
\begin{tabularx}{\textwidth}{L{0.18\textwidth}Y Y Y}
\toprule
\textbf{Instance} & \textbf{Original Query Signal} & \textbf{Rewritten Query} & \textbf{Rewrite Effect} \\
\midrule
\texttt{streamlink-5926}
\newline bug report
& A 1,904-character raw report contains checklist boilerplate, contribution-guideline confirmations, sample Kuwaiti channel URLs, and a debug log ending with ``No plugin can handle URL''.
& ``I'm trying to stream Kuwaiti channels with Streamlink, but it fails with a `No plugin can handle URL' error for the media.gov.kw LiveTV page. I also tried the Drama channel URL. How do I fix this?''
& Reduces checklist and debug boilerplate while preserving the concrete failure, affected site URL, and developer-facing repair request. \\
\midrule
\texttt{fmt-3729}
\newline feature request
& The original describes generic versus native formatting for filesystem paths, including Windows-style examples and generic-path output.
& ``Add support for generic and native format specifiers for filesystem paths to allow cross-platform control over slash direction in output.''
& Converts a verbose feature discussion into a concise search-style request while keeping the target API and desired behavior. \\
\midrule
\texttt{camunda-27226}
\newline refactoring
& The original asks to change the \texttt{state} field from string to enum in ProcessInstance and ProcessInstanceFilter, aligned with OpenAPI values \texttt{ACTIVE}, \texttt{COMPLETED}, and \texttt{CANCELED}.
& ``I'm working with the Camunda Java client and noticed the state field in ProcessInstance and ProcessInstanceFilter is currently just a string. The OpenAPI spec defines it as an enum. How do I update the client code to use an enum type instead?''
& Preserves class names, field name, and enum motivation while turning a specification note into a developer question. \\
\bottomrule
\end{tabularx}
\end{table*}

\section{Detailed \benchname{} Results}
\label{app:detailed-results}

These tables give the detailed results behind the main comparison. The appendix includes BM25 as the sparse-retrieval baseline, while all embedding-model groups are marked as dense retrieval. Each cell reports NDCG@10/Recall@100. Cell shading follows NDCG@10, while best/second-best marks are computed separately for NDCG@10 and Recall@100. LEVEL-1 abbreviations are: Apps = AppsRetrieval, COIR-CSN = COIRCodeSearchNetRetrieval, CFB-ST = CodeFeedbackST, CSN-CC = CodeSearchNetCCRetrieval, CSN = CodeSearchNetRetrieval, CTO-C = CodeTransOceanContest, CTO-DL = CodeTransOceanDL, and Text2SQL = SyntheticText2SQL.

Table~\ref{tab:level1} is mainly a calibration point. It shows that strong performance on conventional code-understanding retrieval does not reliably carry over to requirement-conditioned retrieval: some zero-shot and code-specialized models remain competitive on LEVEL-1, while the in-domain SFT models are not uniformly dominant there. Interestingly, after in-domain SFT, the largest LEVEL-1 drops are concentrated in task formats that are superficially close to natural-language-to-code retrieval, suggesting that PR-based supervision shifts the model toward change-target relevance rather than simply strengthening generic code search.

The central comparison is in Tables~\ref{tab:level2-level3} and~\ref{tab:level2-level3-rewrite}. Moving to repository-local development requests changes the ranking more than it simply lowers the scores. In-domain SFT becomes the most stable signal on both LEVEL-2 and LEVEL-3, whereas rewriting the query is not a guaranteed simplification; for several models, compression removes useful request-specific clues.

The remaining breakdowns show where this pattern is more fragile. Tables~\ref{tab:language-results-full} and~\ref{tab:language-results-level3-full} indicate that the gains are broad but not uniform across languages. Table~\ref{tab:query-intent-results-full} shows that underspecified question and documentation queries remain difficult, and Table~\ref{tab:difficulty-results-full} shows that dense or large repositories are where Recall@100 is most easily lost.

\begin{table*}[t]
\centering
\scriptsize
\caption{\benchname{} LEVEL-1 results.}
\label{tab:level1}
\setlength{\tabcolsep}{1.2pt}
\renewcommand{\arraystretch}{1.06}
\resizebox{\textwidth}{!}{%
\begin{tabular}{@{}l*{12}{c}@{}}
\toprule
 & \multicolumn{12}{c}{\textbf{LEVEL-1}} \\
\cmidrule(lr){2-13}
\textbf{Model} & \rotatebox{65}{\textbf{Apps}} & \rotatebox{65}{\textbf{CLARC}} & \rotatebox{65}{\textbf{COIR-CSN}} & \rotatebox{65}{\textbf{CPRet}} & \rotatebox{65}{\textbf{CFB-ST}} & \rotatebox{65}{\textbf{CSN-CC}} & \rotatebox{65}{\textbf{CSN}} & \rotatebox{65}{\textbf{CTO-C}} & \rotatebox{65}{\textbf{CTO-DL}} & \rotatebox{65}{\textbf{CosQA}} & \rotatebox{65}{\textbf{Text2SQL}} & \rotatebox{65}{\textbf{Avg.}} \\
\cmidrule(lr){1-13}
\rowcolor{black!4}\multicolumn{13}{c}{\textbf{\textit{Sparse Retrieval}}}\\
\cmidrule(lr){1-13}
BM25 & 4.8/14.8 & 9.5/49.6 & 40.9/66.1 & 18.6/31.0 & 68.1/89.5 & 54.0/80.4 & 60.0/90.3 & 47.8/84.2 & 34.4/\textbf{100.0} & 18.8/63.8 & 24.9/67.6 & 34.7/67.0 \\
\cmidrule(lr){1-13}
\rowcolor{black!4}\multicolumn{13}{c}{\textbf{\textit{Small Dense Retrieval Models \hspace{0.35em}(<1B)}}}\\
\cmidrule(lr){1-13}
CodeRankEmbed & 13.5/40.1 & 25.6/59.9 & 73.0/90.8 & 16.8/28.8 & 65.4/91.5 & 64.9/88.8 & 83.8/96.5 & 51.9/81.9 & 34.6/\textbf{100.0} & 30.6/86.6 & 61.7/99.2 & 47.4/78.6 \\
mE5-small & 12.7/39.5 & 15.0/54.8 & 56.4/81.1 & 19.7/30.0 & 66.1/93.1 & 53.0/80.7 & 71.6/94.5 & 55.7/86.0 & 31.5/97.8 & 22.8/74.4 & 42.4/93.5 & 40.6/75.0 \\
mE5-base & 5.9/19.7 & 9.7/45.9 & 37.4/64.4 & 12.3/20.9 & 36.8/67.2 & 26.8/52.4 & 60.6/89.7 & 15.3/44.8 & 30.8/97.8 & 12.7/62.0 & 21.2/54.1 & 24.5/56.3 \\
mE5-large & 19.6/50.8 & 15.8/52.5 & 61.7/84.2 & 24.2/36.6 & 67.5/94.0 & 49.2/77.7 & 78.3/96.0 & 56.6/88.7 & 34.8/\textbf{100.0} & 22.8/78.8 & 48.9/94.3 & 43.6/77.6 \\
bge-large-v1.5 & 5.0/17.7 & 14.9/51.5 & 53.5/81.0 & 14.3/24.1 & 64.8/93.6 & 46.2/75.7 & 81.2/96.2 & 45.1/84.2 & 22.2/88.9 & 30.3/88.2 & 38.5/88.5 & 37.8/71.8 \\
bge-m3 & 1.6/8.6 & 8.4/43.8 & 17.1/29.4 & 13.2/23.0 & 30.4/53.1 & 32.3/57.2 & 41.1/71.0 & 28.4/67.4 & 34.7/\textbf{100.0} & 10.8/47.8 & 43.4/90.1 & 23.8/53.8 \\
gte-multi-base & 7.9/27.3 & 20.9/56.9 & 75.0/90.3 & 21.3/33.8 & 74.3/96.3 & 45.4/71.8 & 87.7/97.2 & 63.0/92.8 & 35.0/\textbf{100.0} & 33.0/89.0 & 39.5/92.3 & 45.7/77.1 \\
jina-emb-v5-small & 46.4/85.6 & 29.9/72.6 & 74.8/93.9 & 40.7/60.1 & 82.9/98.6 & 59.1/83.9 & 88.1/97.2 & 79.6/96.4 & 28.2/98.9 & 38.6/95.4 & 66.3/\underline{99.9} & 57.7/89.3 \\
jina-code-0.5b & 81.6/98.8 & 22.4/67.6 & 75.5/94.3 & 57.0/80.8 & 83.3/99.1 & 63.1/89.3 & 87.7/97.4 & 88.1/\underline{99.1} & \cellcolor{second}\underline{36.6}/\textbf{100.0} & 38.4/\underline{97.0} & 65.1/99.9 & 63.5/93.0 \\
Qwen3-0.6B & 75.0/98.3 & 35.2/77.2 & 82.8/96.5 & 57.6/81.0 & 86.4/99.3 & 86.5/97.9 & 90.5/97.7 & 85.7/\underline{99.1} & 31.9/98.9 & 39.2/96.0 & 64.8/99.7 & 66.9/94.7 \\
\cmidrule(lr){1-13}
\rowcolor{black!4}\multicolumn{13}{c}{\textbf{\textit{Medium Dense Retrieval Models \hspace{0.35em}(1B--2B)}}}\\
\cmidrule(lr){1-13}
gte-Qwen2-1.5B & 4.9/19.0 & 22.1/60.6 & 26.6/51.2 & 12.5/23.4 & 45.7/77.4 & 41.7/71.3 & 64.9/92.8 & 65.7/95.0 & 29.2/\textbf{100.0} & 8.5/46.4 & 23.0/55.8 & 31.3/63.0 \\
jina-code-1.5b & 63.6/93.1 & 18.9/73.0 & 51.8/76.9 & 48.8/62.7 & 79.1/96.4 & 59.6/86.4 & 85.6/97.1 & 70.9/95.5 & \cellcolor{best}\textbf{40.1}/\textbf{100.0} & 33.0/94.6 & 66.2/\textbf{100.0} & 56.2/88.7 \\
F2LLM-v2-1.7B & 71.7/95.3 & 14.2/57.3 & 28.9/43.1 & 41.3/60.1 & 72.8/93.0 & 30.6/49.1 & 38.8/62.9 & 64.6/87.8 & 29.9/93.3 & 15.1/53.6 & 66.2/98.9 & 43.1/72.2 \\
\cmidrule(lr){1-13}
\rowcolor{black!4}\multicolumn{13}{c}{\textbf{\textit{Large Dense Retrieval Models \hspace{0.35em}($\geq$4B)}}}\\
\cmidrule(lr){1-13}
F2LLM-v2-4B & 76.9/94.8 & 30.4/71.8 & 80.3/95.4 & \cellcolor{best}\textbf{80.6}/\textbf{94.0} & 88.2/99.5 & \cellcolor{second}\underline{90.0}/98.2 & 84.2/96.1 & 92.2/98.2 & 32.5/\textbf{100.0} & 30.9/86.6 & \cellcolor{best}\textbf{69.3}/\textbf{100.0} & 68.7/94.0 \\
Qwen3-4B & \cellcolor{second}\underline{89.1}/99.7 & \cellcolor{best}\textbf{48.1}/\textbf{87.4} & 86.3/\underline{97.7} & 73.1/89.3 & 89.5/\underline{99.7} & 87.0/\underline{98.6} & \cellcolor{second}\underline{92.1}/98.0 & \cellcolor{second}\underline{93.7}/\textbf{99.5} & 34.9/\textbf{100.0} & \cellcolor{second}\underline{39.6}/96.6 & 66.5/99.9 & \cellcolor{best}\textbf{72.7}/\textbf{96.9} \\
pplx-embed-4b & 84.8/\textbf{99.9} & 27.2/64.8 & 74.8/93.7 & 52.4/72.5 & 73.6/98.3 & 84.7/97.3 & 87.3/96.8 & 85.1/97.7 & 36.4/\textbf{100.0} & 36.1/95.2 & 67.1/99.8 & 64.5/92.4 \\
C2LLM-7B & 86.0/99.1 & 43.9/85.1 & \cellcolor{second}\underline{88.3}/97.6 & \cellcolor{second}\underline{73.6}/\underline{91.8} & \cellcolor{best}\textbf{90.1}/\textbf{99.7} & \cellcolor{best}\textbf{93.7}/\textbf{99.2} & 91.6/\underline{98.2} & 91.6/\textbf{99.5} & 35.1/98.9 & 37.3/95.0 & 65.4/\textbf{100.0} & \cellcolor{second}\underline{72.4}/96.7 \\
e5-mistral-7b & 26.2/60.0 & 38.1/78.9 & 55.8/81.0 & 41.1/61.7 & 75.8/97.7 & 64.1/90.0 & 84.7/96.9 & 88.0/97.7 & 35.1/\textbf{100.0} & 29.5/87.4 & 63.0/99.5 & 54.7/86.4 \\
SweRankEmbed-L & 15.8/45.2 & 25.9/68.5 & 48.6/72.9 & 26.0/41.7 & 77.1/97.1 & 56.2/84.3 & 77.7/96.3 & 74.9/97.3 & 29.8/\textbf{100.0} & 23.8/77.0 & 43.5/90.4 & 45.4/79.1 \\
F2LLM-v2-8B & 76.1/92.8 & 27.5/75.3 & 66.9/85.6 & 72.1/84.8 & 89.5/99.6 & 77.1/93.1 & 67.4/89.9 & 92.5/98.6 & 32.0/\textbf{100.0} & 33.6/91.6 & \cellcolor{second}\underline{67.7}/\textbf{100.0} & 63.8/91.9 \\
Qwen3-8B & \cellcolor{best}\textbf{91.0}/\underline{99.8} & \cellcolor{second}\underline{48.0}/\underline{87.2} & \cellcolor{best}\textbf{88.5}/\textbf{98.2} & 73.2/89.9 & \cellcolor{second}\underline{89.9}/99.7 & 72.9/95.5 & \cellcolor{best}\textbf{92.4}/\textbf{98.3} & \cellcolor{best}\textbf{94.1}/\textbf{99.5} & 33.2/\textbf{100.0} & \cellcolor{best}\textbf{39.8}/\textbf{97.6} & 65.8/\underline{99.9} & 71.7/\underline{96.9} \\
\cmidrule(lr){1-13}
\rowcolor{black!4}\multicolumn{13}{c}{\textbf{\textit{In-domain SFT Dense Retrieval Models \hspace{0.35em}(0.6B--8B)}}}\\
\cmidrule(lr){1-13}
Qwen3-0.6B-SFT & 65.8/96.0 & 27.6/65.7 & 63.4/88.2 & 54.0/77.2 & 79.5/98.4 & 62.6/89.9 & 83.2/97.4 & 82.7/98.6 & 33.3/98.9 & 30.5/85.6 & 56.8/97.9 & 58.1/90.3 \\
Qwen3-4B-SFT & 82.2/99.1 & 30.4/69.9 & 61.0/87.2 & 63.7/82.8 & 81.3/98.9 & 65.5/91.9 & 82.0/97.6 & 85.6/98.2 & 31.1/98.9 & 22.5/73.4 & 52.2/97.8 & 59.8/90.5 \\
Qwen3-8B-SFT & 82.9/99.4 & 35.6/75.7 & 68.2/91.1 & 66.6/84.9 & 83.6/99.3 & 69.9/93.8 & 85.1/97.6 & 89.5/98.6 & 31.1/\underline{99.4} & 24.4/79.4 & 55.6/98.6 & 63.0/92.5 \\
\bottomrule
\end{tabular}}
\end{table*}

\begin{table*}[t]
\centering
\scriptsize
\caption{\benchname{} LEVEL-2 and LEVEL-3 results. Abbrev.: Pro/Verf/Live/++/+/Multi/Mling = Pro/Verified/Live/++/+/Multi-SWE/Multilingual.}
\label{tab:level2-level3}
\setlength{\tabcolsep}{1.2pt}
\renewcommand{\arraystretch}{1.06}
\resizebox{\textwidth}{!}{%
\begin{tabular}{@{}l*{13}{c}@{}}
\toprule
 & \multicolumn{8}{c}{\textbf{LEVEL-2}} & \multicolumn{5}{c}{\textbf{LEVEL-3}} \\
\cmidrule(lr){2-9} \cmidrule(lr){10-14}
\textbf{Model} & \rotatebox{65}{\textbf{Pro}} & \rotatebox{65}{\textbf{Verf}} & \rotatebox{65}{\textbf{Live}} & \rotatebox{65}{\textbf{++}} & \rotatebox{65}{\textbf{+}} & \rotatebox{65}{\textbf{Multi}} & \rotatebox{65}{\textbf{Mling}} & \rotatebox{65}{\textbf{Avg.}} & \rotatebox{65}{\textbf{Pro}} & \rotatebox{65}{\textbf{Verf}} & \rotatebox{65}{\textbf{Multi}} & \rotatebox{65}{\textbf{Mling}} & \rotatebox{65}{\textbf{Avg.}} \\
\cmidrule(lr){1-14}
\rowcolor{black!4}\multicolumn{14}{c}{\textbf{\textit{Sparse Retrieval}}}\\
\cmidrule(lr){1-14}
BM25 & 19.5/38.0 & 11.9/46.7 & 14.8/44.4 & 12.6/32.2 & 15.0/48.4 & 12.3/37.0 & 5.0/24.8 & 13.0/38.8 & 29.3/36.1 & 22.2/33.9 & 19.7/28.1 & 10.9/22.2 & 20.5/30.1 \\
\cmidrule(lr){1-14}
\rowcolor{black!4}\multicolumn{14}{c}{\textbf{\textit{Small Dense Retrieval Models \hspace{0.35em}(<1B)}}}\\
\cmidrule(lr){1-14}
CodeRankEmbed & 16.9/34.4 & 12.0/37.7 & 12.2/37.9 & 11.6/29.5 & 11.3/33.9 & 14.7/38.1 & 5.7/18.7 & 12.1/32.9 & 27.0/32.8 & 20.6/30.6 & 28.3/30.6 & 14.2/20.3 & 22.5/28.6 \\
mE5-small & 21.2/41.1 & 7.7/43.4 & 11.0/39.6 & 11.7/31.5 & 9.8/38.3 & 12.4/33.6 & 4.6/18.6 & 11.2/35.2 & 30.3/38.4 & 17.6/31.0 & 23.9/29.3 & 9.1/19.0 & 20.2/29.4 \\
mE5-base & 6.0/14.8 & 1.9/11.7 & 5.2/25.2 & 4.7/17.7 & 4.1/19.0 & 2.9/13.2 & 2.2/8.8 & 3.9/15.8 & 8.8/13.9 & 8.1/15.6 & 7.9/13.1 & 3.2/9.3 & 7.0/13.0 \\
mE5-large & 17.0/35.8 & 9.0/45.5 & 11.0/39.2 & 11.1/29.8 & 10.1/38.6 & 10.5/34.5 & 3.6/21.6 & 10.3/35.0 & 24.2/32.2 & 21.3/26.8 & 20.4/26.8 & 10.7/20.0 & 19.2/26.5 \\
bge-large-v1.5 & 19.1/35.1 & 11.9/47.1 & 13.9/45.7 & 14.0/33.1 & 16.7/44.0 & 11.9/41.0 & 4.6/24.2 & 13.1/38.6 & 26.4/33.1 & 25.0/34.6 & 24.5/30.5 & 10.8/21.5 & 21.7/29.9 \\
bge-m3 & 6.3/15.1 & 2.5/18.0 & 4.5/22.8 & 5.0/18.9 & 6.5/28.2 & 5.1/14.6 & 2.4/10.5 & 4.6/18.3 & 8.9/12.9 & 7.2/11.8 & 13.1/14.9 & 5.5/10.2 & 8.7/12.5 \\
gte-multi-base & 15.2/30.2 & 9.4/34.4 & 13.7/44.6 & 12.6/31.5 & 16.8/38.1 & 12.9/37.1 & 5.0/24.1 & 12.2/34.3 & 22.5/29.0 & 22.2/25.4 & 23.9/28.8 & 12.9/23.0 & 20.4/26.6 \\
jina-emb-v5-small & 28.8/52.1 & 19.0/53.4 & 18.2/51.8 & 18.2/40.7 & 19.1/48.4 & 18.7/47.4 & 11.4/33.5 & 19.1/46.7 & 42.1/50.3 & 34.0/39.9 & 33.8/39.7 & 22.0/34.3 & 33.0/41.1 \\
jina-code-0.5b & 29.8/51.7 & 17.9/63.7 & 20.7/56.3 & 19.8/43.9 & 22.6/52.0 & 22.0/54.1 & 12.8/41.6 & 20.8/51.9 & 42.0/50.2 & 37.6/48.6 & 41.0/42.9 & 26.1/40.2 & 36.7/45.5 \\
Qwen3-0.6B & 24.0/48.9 & 15.7/57.2 & 17.2/50.4 & 15.9/37.4 & 22.9/49.4 & 16.3/48.2 & 7.1/27.2 & 17.0/45.5 & 37.3/46.7 & 37.2/45.4 & 35.5/38.3 & 20.2/30.2 & 32.6/40.2 \\
\cmidrule(lr){1-14}
\rowcolor{black!4}\multicolumn{14}{c}{\textbf{\textit{Medium Dense Retrieval Models \hspace{0.35em}(1B--2B)}}}\\
\cmidrule(lr){1-14}
gte-Qwen2-1.5B & 6.4/19.6 & 0.7/13.9 & 3.4/17.8 & 4.8/18.7 & 2.3/10.1 & 4.3/21.0 & 2.5/10.3 & 3.5/15.9 & 10.0/18.3 & 8.2/12.9 & 8.3/17.5 & 5.9/10.8 & 8.1/14.9 \\
jina-code-1.5b & 25.5/50.4 & 11.0/55.4 & 17.8/55.7 & 17.0/41.0 & 20.5/52.9 & 17.1/47.1 & 10.0/37.0 & 17.0/48.5 & 39.0/49.0 & 32.3/45.6 & 34.1/40.4 & 21.2/35.7 & 31.6/42.7 \\
F2LLM-v2-1.7B & 4.5/11.9 & 8.6/28.5 & 7.3/32.1 & 6.5/20.4 & 7.3/25.1 & 5.4/18.8 & 1.4/8.3 & 5.8/20.7 & 7.1/11.0 & 8.0/17.1 & 12.2/15.5 & 1.8/6.4 & 7.3/12.5 \\
\cmidrule(lr){1-14}
\rowcolor{black!4}\multicolumn{14}{c}{\textbf{\textit{Large Dense Retrieval Models \hspace{0.35em}($\geq$4B)}}}\\
\cmidrule(lr){1-14}
F2LLM-v2-4B & 27.1/50.4 & 20.5/57.4 & 19.7/56.9 & 17.3/38.7 & 21.0/56.1 & 21.6/49.6 & 10.9/40.0 & 19.7/49.9 & 40.3/47.8 & 37.6/43.3 & 39.3/38.9 & 24.2/36.2 & 35.3/41.6 \\
Qwen3-4B & 24.7/49.8 & 22.3/59.1 & 18.2/51.1 & 17.2/38.4 & 21.7/52.9 & 16.9/46.3 & 7.3/30.8 & 18.3/46.9 & 36.5/48.2 & 40.9/47.0 & 36.2/38.9 & 17.5/29.1 & 32.8/40.8 \\
pplx-embed-4b & 19.4/39.3 & 10.7/45.4 & 13.4/41.5 & 12.0/31.9 & 16.0/42.8 & 13.1/35.2 & 5.4/22.1 & 12.9/36.9 & 30.1/38.4 & 30.4/37.6 & 27.5/30.3 & 12.3/21.8 & 25.0/32.1 \\
C2LLM-7B & 25.9/51.2 & 13.0/52.9 & 17.4/48.8 & 18.1/39.6 & 15.5/44.1 & 17.7/45.4 & 9.2/32.2 & 16.7/44.9 & 40.7/50.1 & 31.3/40.1 & 38.5/38.5 & 21.3/35.3 & 32.9/41.0 \\
e5-mistral-7b & 24.9/50.7 & 16.1/61.9 & 19.9/54.3 & 18.2/43.9 & 21.2/56.8 & 17.9/51.6 & 10.1/42.6 & 18.3/51.7 & 37.8/48.6 & 39.1/48.0 & 33.7/40.6 & 24.9/40.6 & 33.9/44.4 \\
SweRankEmbed-L & 31.3/52.9 & 25.1/66.1 & 25.0/57.6 & 19.5/42.6 & 24.5/57.5 & 20.0/47.8 & 11.2/40.2 & 22.4/52.1 & 44.3/50.9 & 37.0/46.9 & 35.9/37.4 & 20.4/34.3 & 34.4/42.4 \\
F2LLM-v2-8B & 21.6/37.9 & 24.7/57.5 & 18.9/53.3 & 17.0/37.0 & 24.8/55.7 & 15.6/42.4 & 10.6/36.9 & 19.0/45.8 & 34.2/38.1 & 42.4/46.2 & 33.1/32.2 & 25.5/32.3 & 33.8/37.2 \\
Qwen3-8B & 23.5/44.7 & 21.4/59.2 & 19.8/51.4 & 18.6/40.0 & 27.6/52.7 & 20.2/49.6 & 11.1/38.3 & 20.3/48.0 & 34.0/43.0 & 42.6/48.1 & 37.6/39.5 & 23.6/35.5 & 34.4/41.5 \\
\cmidrule(lr){1-14}
\rowcolor{black!4}\multicolumn{14}{c}{\textbf{\textit{In-domain SFT Dense Retrieval Models \hspace{0.35em}(0.6B--8B)}}}\\
\cmidrule(lr){1-14}
Qwen3-0.6B-SFT & 36.7/61.2 & 25.0/68.9 & 25.7/62.8 & 23.7/49.2 & 29.2/60.0 & 28.5/63.7 & 16.6/50.1 & 26.5/59.4 & 52.6/59.3 & 44.3/57.7 & 46.9/51.6 & 34.3/49.0 & 44.5/54.4 \\
Qwen3-4B-SFT & \cellcolor{second}\underline{37.9}/\textbf{65.2} & \cellcolor{second}\underline{31.1}/\textbf{79.5} & \cellcolor{second}\underline{28.3}/\underline{66.3} & \cellcolor{second}\underline{27.7}/\underline{56.2} & \cellcolor{second}\underline{32.7}/\underline{65.0} & \cellcolor{second}\underline{30.5}/\underline{65.7} & \cellcolor{best}\textbf{24.0}/\underline{64.1} & \cellcolor{second}\underline{30.3}/\underline{66.0} & \cellcolor{second}\underline{54.8}/\underline{64.8} & \cellcolor{second}\underline{49.6}/\textbf{66.2} & \cellcolor{second}\underline{51.6}/\underline{54.2} & \cellcolor{second}\underline{41.0}/\underline{61.0} & \cellcolor{second}\underline{49.2}/\textbf{61.6} \\
Qwen3-8B-SFT & \cellcolor{best}\textbf{39.3}/\underline{64.9} & \cellcolor{best}\textbf{33.9}/\underline{76.1} & \cellcolor{best}\textbf{33.1}/\textbf{68.0} & \cellcolor{best}\textbf{29.6}/\textbf{57.1} & \cellcolor{best}\textbf{36.0}/\textbf{66.0} & \cellcolor{best}\textbf{34.0}/\textbf{67.5} & \cellcolor{second}\underline{23.7}/\textbf{65.2} & \cellcolor{best}\textbf{32.8}/\textbf{66.4} & \cellcolor{best}\textbf{56.2}/\textbf{64.9} & \cellcolor{best}\textbf{51.1}/\underline{63.3} & \cellcolor{best}\textbf{52.1}/\textbf{55.3} & \cellcolor{best}\textbf{41.5}/\textbf{62.0} & \cellcolor{best}\textbf{50.2}/\underline{61.4} \\
\bottomrule
\end{tabular}}
\end{table*}

\begin{table*}[t]
\centering
\scriptsize
\caption{\benchname{} rewritten-query results. Abbrev.: Pro/Verf/Live/++/+/Multi/Mling = Pro/Verified/Live/++/+/Multi-SWE/Multilingual.}
\label{tab:level2-level3-rewrite}
\setlength{\tabcolsep}{1.2pt}
\renewcommand{\arraystretch}{1.06}
\resizebox{\textwidth}{!}{%
\begin{tabular}{@{}l*{13}{c}@{}}
\toprule
 & \multicolumn{8}{c}{\textbf{LEVEL-2-Rw}} & \multicolumn{5}{c}{\textbf{LEVEL-3-Rw}} \\
\cmidrule(lr){2-9} \cmidrule(lr){10-14}
\textbf{Model} & \rotatebox{65}{\textbf{Pro}} & \rotatebox{65}{\textbf{Verf}} & \rotatebox{65}{\textbf{Live}} & \rotatebox{65}{\textbf{++}} & \rotatebox{65}{\textbf{+}} & \rotatebox{65}{\textbf{Multi}} & \rotatebox{65}{\textbf{Mling}} & \rotatebox{65}{\textbf{Avg.}} & \rotatebox{65}{\textbf{Pro}} & \rotatebox{65}{\textbf{Verf}} & \rotatebox{65}{\textbf{Multi}} & \rotatebox{65}{\textbf{Mling}} & \rotatebox{65}{\textbf{Avg.}} \\
\cmidrule(lr){1-14}
\rowcolor{black!4}\multicolumn{14}{c}{\textbf{\textit{Sparse Retrieval}}}\\
\cmidrule(lr){1-14}
BM25 & 18.2/35.9 & 13.6/47.9 & 18.0/47.6 & 14.0/33.9 & 17.3/50.8 & 12.0/39.7 & 3.3/19.3 & 13.8/39.3 & 26.3/34.7 & 23.6/33.6 & 22.0/28.3 & 7.2/18.0 & 19.8/28.7 \\
\cmidrule(lr){1-14}
\rowcolor{black!4}\multicolumn{14}{c}{\textbf{\textit{Small Dense Retrieval Models \hspace{0.35em}(<1B)}}}\\
\cmidrule(lr){1-14}
CodeRankEmbed & 20.4/40.3 & 13.7/50.0 & 14.2/42.2 & 12.4/32.3 & 15.6/38.9 & 15.9/41.5 & 4.0/20.2 & 13.8/37.9 & 30.7/37.5 & 30.1/36.5 & 33.2/32.9 & 11.0/18.5 & 26.2/31.4 \\
mE5-small & 18.8/38.2 & 10.0/42.9 & 12.2/40.1 & 11.5/32.0 & 11.7/39.8 & 13.3/40.2 & 2.6/15.1 & 11.4/35.5 & 28.5/36.6 & 19.5/32.1 & 25.6/31.2 & 6.0/14.1 & 19.9/28.5 \\
mE5-base & 3.5/11.3 & 1.4/12.9 & 4.7/25.6 & 4.1/17.4 & 2.9/18.9 & 1.6/10.3 & 0.2/3.1 & 2.6/14.2 & 6.3/12.4 & 7.3/14.0 & 5.9/12.2 & 1.3/4.3 & 5.2/10.7 \\
mE5-large & 14.8/33.0 & 11.3/46.0 & 9.7/38.6 & 10.3/29.9 & 10.5/39.4 & 13.2/37.4 & 1.7/13.7 & 10.2/34.0 & 22.4/30.5 & 21.5/28.8 & 22.1/28.1 & 5.9/13.5 & 18.0/25.2 \\
bge-large-v1.5 & 18.7/35.4 & 10.0/44.7 & 14.2/45.8 & 12.3/31.5 & 13.8/40.3 & 13.0/43.0 & 1.9/15.2 & 12.0/36.6 & 26.1/33.1 & 20.1/32.2 & 28.1/31.3 & 5.3/12.8 & 19.9/27.4 \\
bge-m3 & 4.8/11.1 & 2.9/17.6 & 3.9/18.9 & 4.4/15.5 & 5.2/20.8 & 5.0/14.9 & 1.0/8.0 & 3.9/15.3 & 7.6/10.6 & 9.6/14.0 & 10.9/14.3 & 3.1/6.7 & 7.8/11.4 \\
gte-multi-base & 18.2/32.8 & 10.9/41.2 & 16.6/48.6 & 13.8/33.1 & 16.7/38.9 & 14.7/40.4 & 3.7/26.1 & 13.5/37.3 & 27.1/32.2 & 29.1/30.7 & 28.4/30.9 & 9.3/21.4 & 23.5/28.8 \\
jina-emb-v5-small & 28.3/50.8 & 12.5/57.8 & 17.9/51.6 & 17.0/39.6 & 16.9/47.5 & 20.4/51.1 & 3.6/25.7 & 16.7/46.3 & 39.3/49.5 & 35.7/42.4 & 38.2/41.8 & 11.1/23.7 & 31.1/39.3 \\
jina-code-0.5b & 30.4/52.5 & 11.8/55.9 & 18.7/52.6 & 19.6/43.5 & 18.1/46.7 & 21.1/54.1 & 7.1/27.0 & 18.1/47.5 & 43.2/51.9 & 35.1/44.0 & 39.5/44.0 & 13.8/29.0 & 32.9/42.2 \\
Qwen3-0.6B & 25.0/49.0 & 14.4/61.8 & 17.7/52.4 & 16.3/38.3 & 21.2/48.0 & 17.5/49.3 & 3.5/19.3 & 16.5/45.5 & 38.5/47.0 & 36.7/46.3 & 37.0/40.6 & 12.4/23.1 & 31.1/39.2 \\
\cmidrule(lr){1-14}
\rowcolor{black!4}\multicolumn{14}{c}{\textbf{\textit{Medium Dense Retrieval Models \hspace{0.35em}(1B--2B)}}}\\
\cmidrule(lr){1-14}
gte-Qwen2-1.5B & 7.7/21.4 & 0.4/14.5 & 3.6/19.5 & 6.2/22.3 & 2.4/13.6 & 9.2/26.8 & 0.8/9.2 & 4.3/18.2 & 11.4/20.1 & 6.6/13.2 & 16.8/21.9 & 3.3/7.7 & 9.5/15.7 \\
jina-code-1.5b & 24.4/50.6 & 9.1/52.9 & 15.7/53.4 & 16.2/41.2 & 17.6/49.2 & 17.4/51.2 & 3.5/26.6 & 14.9/46.4 & 38.5/48.9 & 31.3/43.3 & 36.2/41.4 & 9.6/23.1 & 28.9/39.2 \\
F2LLM-v2-1.7B & 5.7/14.9 & 6.7/23.7 & 7.3/32.9 & 6.2/20.6 & 6.4/23.4 & 5.7/18.9 & 0.6/3.9 & 5.5/19.8 & 10.1/14.1 & 8.5/14.6 & 11.2/14.5 & 0.7/2.0 & 7.6/11.3 \\
\cmidrule(lr){1-14}
\rowcolor{black!4}\multicolumn{14}{c}{\textbf{\textit{Large Dense Retrieval Models \hspace{0.35em}($\geq$4B)}}}\\
\cmidrule(lr){1-14}
F2LLM-v2-4B & 26.0/50.0 & 19.9/58.0 & 18.7/56.4 & 17.0/38.8 & 19.0/52.7 & 21.7/50.2 & 5.4/25.3 & 18.3/47.4 & 39.5/47.6 & 37.6/43.5 & 39.0/37.9 & 13.2/25.2 & 32.3/38.6 \\
Qwen3-4B & 24.0/48.2 & 19.9/59.4 & 17.9/52.6 & 16.7/40.3 & 18.9/52.3 & 19.2/48.1 & 4.2/22.5 & 17.3/46.2 & 36.1/47.4 & 39.7/47.4 & 38.6/40.7 & 10.1/22.0 & 31.1/39.3 \\
pplx-embed-4b & 19.5/39.7 & 8.7/48.6 & 12.5/41.6 & 11.3/32.5 & 14.1/43.2 & 12.0/33.8 & 2.1/16.4 & 11.5/36.5 & 30.0/38.2 & 28.1/36.5 & 25.9/28.3 & 6.8/15.8 & 22.7/29.7 \\
C2LLM-7B & 25.0/48.7 & 11.7/50.4 & 15.2/48.3 & 17.1/38.6 & 14.2/40.7 & 14.1/39.6 & 4.2/23.4 & 14.5/41.4 & 36.7/47.7 & 30.7/40.4 & 34.3/36.9 & 11.9/26.0 & 28.4/37.8 \\
e5-mistral-7b & 24.8/50.8 & 15.3/61.2 & 19.3/53.6 & 17.7/43.7 & 20.7/53.4 & 20.0/52.8 & 3.9/33.6 & 17.4/49.9 & 38.3/49.3 & 39.1/46.2 & 38.1/42.9 & 11.6/29.3 & 31.8/42.0 \\
SweRankEmbed-L & 29.9/51.2 & 25.6/65.2 & 23.4/57.0 & 20.3/41.8 & 26.1/52.2 & 21.0/50.3 & 5.8/30.2 & 21.7/49.7 & 43.8/50.3 & 41.9/51.8 & 36.3/38.7 & 12.5/25.4 & 33.6/41.5 \\
F2LLM-v2-8B & 20.5/37.0 & 24.6/53.0 & 18.6/53.6 & 16.6/36.9 & 22.2/54.7 & 15.7/43.3 & 6.1/24.8 & 17.8/43.3 & 33.6/37.3 & 40.8/47.1 & 33.6/33.0 & 15.3/22.2 & 30.8/34.9 \\
Qwen3-8B & 23.8/45.0 & 16.0/57.9 & 17.7/50.0 & 16.1/39.5 & 22.2/49.4 & 19.3/49.6 & 5.1/24.2 & 17.2/45.1 & 33.8/42.1 & 39.1/49.3 & 37.7/39.6 & 12.8/22.4 & 30.8/38.4 \\
\cmidrule(lr){1-14}
\rowcolor{black!4}\multicolumn{14}{c}{\textbf{\textit{In-domain SFT Dense Retrieval Models \hspace{0.35em}(0.6B--8B)}}}\\
\cmidrule(lr){1-14}
Qwen3-0.6B-SFT & 35.6/59.1 & 28.0/73.4 & 25.0/60.9 & 23.0/49.6 & 28.0/61.0 & \cellcolor{second}\underline{30.2}/63.4 & 8.6/43.0 & 25.5/58.6 & 51.2/58.3 & 43.2/60.5 & 47.0/50.1 & 19.5/39.2 & 40.2/52.0 \\
Qwen3-4B-SFT & \cellcolor{second}\underline{37.0}/\textbf{63.8} & \cellcolor{second}\underline{30.0}/\textbf{78.8} & \cellcolor{second}\underline{27.9}/\underline{66.5} & \cellcolor{second}\underline{25.8}/\underline{56.5} & \cellcolor{second}\underline{28.2}/\underline{65.2} & 29.3/\underline{64.9} & \cellcolor{second}\underline{12.7}/\underline{54.1} & \cellcolor{second}\underline{27.3}/\textbf{64.3} & \cellcolor{second}\underline{54.3}/\underline{63.9} & \cellcolor{second}\underline{48.5}/\textbf{66.1} & \cellcolor{best}\textbf{50.9}/\textbf{53.7} & \cellcolor{best}\textbf{22.4}/\underline{48.5} & \cellcolor{second}\underline{44.1}/\textbf{58.1} \\
Qwen3-8B-SFT & \cellcolor{best}\textbf{39.6}/\underline{63.8} & \cellcolor{best}\textbf{30.4}/\underline{75.2} & \cellcolor{best}\textbf{31.3}/\textbf{66.8} & \cellcolor{best}\textbf{27.8}/\textbf{56.7} & \cellcolor{best}\textbf{31.1}/\textbf{65.4} & \cellcolor{best}\textbf{31.0}/\textbf{65.8} & \cellcolor{best}\textbf{13.4}/\textbf{54.7} & \cellcolor{best}\textbf{29.2}/\underline{64.0} & \cellcolor{best}\textbf{56.1}/\textbf{64.3} & \cellcolor{best}\textbf{48.6}/\underline{62.7} & \cellcolor{second}\underline{50.8}/\underline{53.5} & \cellcolor{second}\underline{22.2}/\textbf{49.3} & \cellcolor{best}\textbf{44.4}/\underline{57.4} \\
\bottomrule
\end{tabular}}
\end{table*}

\begin{table*}[t]
\centering
\scriptsize
\caption{LEVEL-2 per-language results.}
\label{tab:language-results-full}
\setlength{\tabcolsep}{1.2pt}
\renewcommand{\arraystretch}{1.06}
\resizebox{\textwidth}{!}{%
\begin{tabular}{@{}l*{11}{c}@{}}
\toprule
 & \multicolumn{11}{c}{\textbf{LEVEL-2}} \\
\cmidrule(lr){2-12}
\textbf{Model} & \rotatebox{65}{\textbf{Python}} & \rotatebox{65}{\textbf{Go}} & \rotatebox{65}{\textbf{JS}} & \rotatebox{65}{\textbf{Rust}} & \rotatebox{65}{\textbf{TS}} & \rotatebox{65}{\textbf{Java}} & \rotatebox{65}{\textbf{C++}} & \rotatebox{65}{\textbf{C}} & \rotatebox{65}{\textbf{Ruby}} & \rotatebox{65}{\textbf{PHP}} & \rotatebox{65}{\textbf{Swift}} \\
\cmidrule(lr){1-12}
\rowcolor{black!4}\multicolumn{12}{c}{\textbf{\textit{Sparse Retrieval}}}\\
\cmidrule(lr){1-12}
BM25 & 15.3/41.2 & 16.0/38.2 & 8.7/32.4 & 11.5/36.3 & 7.4/24.3 & 8.1/20.6 & 9.7/22.5 & 11.9/24.3 & 6.4/23.8 & 5.8/17.3 & 0.0/0.0 \\
\cmidrule(lr){1-12}
\rowcolor{black!4}\multicolumn{12}{c}{\textbf{\textit{Small Dense Retrieval Models \hspace{0.35em}(<1B)}}}\\
\cmidrule(lr){1-12}
CodeRankEmbed & 10.7/32.0 & 15.3/37.6 & 8.8/27.2 & 12.5/35.9 & 10.2/28.4 & 8.7/22.8 & 14.1/28.0 & 12.4/32.2 & 5.4/14.0 & 6.4/14.1 & 0.0/0.0 \\
mE5-small & 11.6/36.3 & 14.1/33.6 & 9.4/31.7 & 10.5/31.4 & 9.3/23.8 & 7.4/19.6 & 11.7/26.0 & 7.4/21.8 & 5.5/14.9 & 7.1/21.6 & 0.0/0.0 \\
mE5-base & 4.2/18.2 & 3.4/9.8 & 2.9/14.9 & 3.5/12.7 & 2.9/9.6 & 3.0/9.0 & 3.6/7.1 & 1.3/7.1 & 0.6/4.4 & 2.3/6.0 & 0.0/0.0 \\
mE5-large & 11.3/36.5 & 12.1/30.8 & 8.9/33.2 & 8.7/26.7 & 9.4/27.2 & 7.8/18.2 & 9.1/18.8 & 6.1/21.5 & 4.6/18.1 & 5.4/19.1 & 0.0/0.0 \\
bge-large-v1.5 & 15.5/42.1 & 11.5/26.7 & 12.0/39.5 & 10.2/33.1 & 9.1/28.7 & 9.1/21.6 & 13.5/28.6 & 9.8/31.2 & 4.4/17.0 & 3.6/19.6 & 0.0/\textbf{100.0} \\
bge-m3 & 5.1/19.1 & 4.5/13.9 & 4.0/17.7 & 4.7/13.3 & 2.2/7.8 & 3.0/8.1 & 3.8/11.2 & 3.0/11.9 & 2.6/5.7 & 2.1/11.4 & 0.0/0.0 \\
gte-multi-base & 13.9/37.9 & 10.8/25.7 & 9.1/29.1 & 11.6/36.0 & 13.3/35.2 & 10.2/26.1 & 12.5/29.7 & 7.8/25.3 & 5.2/17.0 & 6.5/22.9 & 0.0/0.0 \\
jina-emb-v5-small & 18.5/49.1 & 24.5/51.3 & 16.0/45.8 & 17.8/45.0 & 15.3/45.5 & 11.6/28.7 & 17.9/34.3 & 10.8/33.7 & 7.1/29.0 & 12.8/24.0 & \cellcolor{second}\underline{85.0}/\textbf{100.0} \\
jina-code-0.5b & 22.0/54.2 & 26.0/54.4 & 21.1/54.1 & 22.2/50.3 & 24.3/55.0 & 14.7/35.7 & 20.9/41.1 & 13.2/35.9 & 10.2/39.1 & 11.4/34.7 & 21.8/\underline{50.0} \\
Qwen3-0.6B & 18.4/49.3 & 16.5/43.7 & 18.1/50.0 & 14.3/37.7 & 17.0/50.7 & 10.2/27.2 & 17.7/33.8 & 11.8/34.7 & 8.0/24.9 & 4.5/15.8 & 17.7/\textbf{100.0} \\
\cmidrule(lr){1-12}
\rowcolor{black!4}\multicolumn{12}{c}{\textbf{\textit{Medium Dense Retrieval Models \hspace{0.35em}(1B--2B)}}}\\
\cmidrule(lr){1-12}
gte-Qwen2-1.5B & 2.0/10.8 & 4.0/15.2 & 3.5/17.0 & 4.1/17.3 & 3.3/12.2 & 3.1/11.9 & 1.8/10.5 & 2.1/14.4 & 0.0/3.7 & 0.9/3.1 & 0.0/0.0 \\
jina-code-1.5b & 18.3/53.0 & 23.1/53.4 & 14.1/46.1 & 18.2/45.5 & 15.9/47.0 & 14.6/30.5 & 16.2/31.3 & 10.5/30.7 & 12.2/38.6 & 9.6/30.5 & 0.0/0.0 \\
F2LLM-v2-1.7B & 6.3/24.6 & 3.1/8.6 & 3.7/14.1 & 3.3/14.0 & 2.2/7.7 & 4.9/13.2 & 3.6/10.6 & 2.2/9.7 & 0.0/3.3 & 0.5/4.7 & 0.0/0.0 \\
\cmidrule(lr){1-12}
\rowcolor{black!4}\multicolumn{12}{c}{\textbf{\textit{Large Dense Retrieval Models \hspace{0.35em}($\geq$4B)}}}\\
\cmidrule(lr){1-12}
F2LLM-v2-4B & 20.9/52.2 & 21.5/50.1 & 23.4/54.3 & 20.0/45.1 & 22.7/50.1 & 12.6/32.6 & 17.8/36.3 & 11.9/35.8 & 8.7/44.3 & 7.5/23.8 & 17.7/\textbf{100.0} \\
Qwen3-4B & 19.3/49.8 & 15.1/42.4 & 16.0/49.4 & 15.6/42.5 & 19.3/49.0 & 10.0/25.2 & 18.4/37.5 & 12.5/37.0 & 4.2/20.6 & 4.4/14.1 & 0.0/\textbf{100.0} \\
pplx-embed-4b & 14.2/40.6 & 16.3/37.2 & 11.7/36.8 & 13.0/35.3 & 12.2/34.3 & 8.2/19.2 & 12.5/27.7 & 7.4/23.5 & 3.2/12.9 & 1.7/14.9 & 0.0/\underline{50.0} \\
C2LLM-7B & 16.6/45.8 & 24.1/52.4 & 20.1/50.7 & 19.4/41.8 & 19.1/48.2 & 12.1/28.4 & 17.3/36.0 & 10.3/33.7 & 11.3/38.3 & 2.2/23.8 & 38.9/\textbf{100.0} \\
e5-mistral-7b & 20.1/54.0 & 25.5/53.5 & 15.2/47.5 & 16.3/44.0 & 19.0/53.5 & 11.3/31.7 & 16.2/39.8 & 14.5/40.1 & 9.6/44.2 & 6.6/38.4 & 0.0/\textbf{100.0} \\
SweRankEmbed-L & 25.9/54.8 & 26.8/51.0 & 21.2/49.9 & 18.0/47.4 & 25.0/52.6 & 10.6/27.0 & 17.9/39.6 & 8.7/30.5 & 9.2/28.5 & \cellcolor{second}\underline{18.7}/38.8 & 0.0/\textbf{100.0} \\
F2LLM-v2-8B & 20.5/50.4 & 17.1/35.7 & 14.7/37.3 & 19.0/41.4 & 17.6/35.9 & 13.2/34.2 & 16.4/37.1 & 13.4/36.9 & 10.6/21.1 & 10.6/39.5 & 38.7/\underline{50.0} \\
Qwen3-8B & 21.9/50.9 & 14.6/42.6 & 18.6/48.8 & 22.1/52.2 & 22.6/48.7 & 12.5/26.9 & 21.0/43.0 & 17.0/40.3 & 13.4/40.3 & 4.2/24.5 & 45.6/\textbf{100.0} \\
\cmidrule(lr){1-12}
\rowcolor{black!4}\multicolumn{12}{c}{\textbf{\textit{In-domain SFT Dense Retrieval Models \hspace{0.35em}(0.6B--8B)}}}\\
\cmidrule(lr){1-12}
Qwen3-0.6B-SFT & 27.5/61.3 & 33.3/62.2 & 31.8/67.8 & 28.1/56.6 & 32.3/66.6 & 17.6/41.3 & 25.4/47.6 & 19.0/44.7 & 12.2/47.3 & 15.5/50.2 & 48.2/\textbf{100.0} \\
Qwen3-4B-SFT & \cellcolor{second}\underline{30.6}/\textbf{67.8} & \cellcolor{second}\underline{36.3}/\underline{67.4} & \cellcolor{second}\underline{36.6}/\textbf{75.2} & \cellcolor{second}\underline{33.4}/\underline{64.1} & \cellcolor{second}\underline{36.0}/\textbf{72.9} & \cellcolor{second}\underline{19.7}/\underline{46.0} & \cellcolor{second}\underline{26.5}/\underline{51.4} & \cellcolor{second}\underline{22.7}/\underline{51.8} & \cellcolor{second}\underline{26.5}/\textbf{65.9} & 14.9/\underline{55.1} & \cellcolor{best}\textbf{92.0}/\textbf{100.0} \\
Qwen3-8B-SFT & \cellcolor{best}\textbf{33.7}/\underline{67.4} & \cellcolor{best}\textbf{38.8}/\textbf{69.6} & \cellcolor{best}\textbf{39.8}/\underline{74.6} & \cellcolor{best}\textbf{33.6}/\textbf{65.5} & \cellcolor{best}\textbf{36.8}/\underline{67.7} & \cellcolor{best}\textbf{21.8}/\textbf{49.5} & \cellcolor{best}\textbf{29.1}/\textbf{53.0} & \cellcolor{best}\textbf{27.8}/\textbf{55.3} & \cellcolor{best}\textbf{27.7}/\underline{64.1} & \cellcolor{best}\textbf{19.4}/\textbf{70.3} & 60.5/\textbf{100.0} \\
\bottomrule
\end{tabular}}
\end{table*}

\begin{table*}[t]
\centering
\scriptsize
\caption{LEVEL-3 per-language results.}
\label{tab:language-results-level3-full}
\setlength{\tabcolsep}{1.2pt}
\renewcommand{\arraystretch}{1.06}
\resizebox{\textwidth}{!}{%
\begin{tabular}{@{}l*{11}{c}@{}}
\toprule
 & \multicolumn{11}{c}{\textbf{LEVEL-3}} \\
\cmidrule(lr){2-12}
\textbf{Model} & \rotatebox{65}{\textbf{Python}} & \rotatebox{65}{\textbf{Go}} & \rotatebox{65}{\textbf{JS}} & \rotatebox{65}{\textbf{Rust}} & \rotatebox{65}{\textbf{TS}} & \rotatebox{65}{\textbf{Java}} & \rotatebox{65}{\textbf{C++}} & \rotatebox{65}{\textbf{C}} & \rotatebox{65}{\textbf{Ruby}} & \rotatebox{65}{\textbf{PHP}} & \rotatebox{65}{\textbf{Swift}} \\
\cmidrule(lr){1-12}
\rowcolor{black!4}\multicolumn{12}{c}{\textbf{\textit{Sparse Retrieval}}}\\
\cmidrule(lr){1-12}
BM25 & 26.2/32.5 & 29.2/30.3 & 10.6/25.9 & 17.9/25.0 & 9.4/14.2 & 17.9/12.4 & 16.8/15.6 & 19.1/15.2 & 11.3/21.4 & 7.4/12.7 & 0.0/0.0 \\
\cmidrule(lr){1-12}
\rowcolor{black!4}\multicolumn{12}{c}{\textbf{\textit{Small Dense Retrieval Models \hspace{0.35em}(<1B)}}}\\
\cmidrule(lr){1-12}
CodeRankEmbed & 26.4/30.7 & 33.5/31.6 & 12.3/23.2 & 26.5/28.6 & 17.6/23.8 & 21.7/13.8 & 28.6/21.9 & 26.3/20.4 & 13.6/19.1 & 10.6/13.6 & 0.0/11.1 \\
mE5-small & 24.2/31.2 & 27.9/28.2 & 13.3/27.6 & 21.9/25.3 & 14.3/20.3 & 18.7/12.2 & 26.7/20.5 & 18.8/15.8 & 10.8/21.0 & 7.9/21.2 & 0.0/0.0 \\
mE5-base & 8.2/12.2 & 8.0/9.7 & 3.6/13.7 & 8.9/11.7 & 3.7/7.6 & 6.3/4.4 & 8.8/6.4 & 5.9/6.2 & 3.7/7.7 & 4.3/5.8 & 0.0/0.0 \\
mE5-large & 21.9/28.2 & 23.7/24.1 & 11.0/26.6 & 18.6/22.6 & 12.5/20.6 & 19.0/11.5 & 17.4/14.9 & 15.6/14.2 & 15.0/20.8 & 7.5/17.5 & 0.0/0.0 \\
bge-large-v1.5 & 26.8/32.6 & 23.0/21.5 & 16.4/34.4 & 24.9/27.5 & 13.8/21.7 & 18.3/11.8 & 26.8/22.1 & 21.9/19.8 & 12.9/21.5 & 6.0/18.0 & 0.0/22.2 \\
bge-m3 & 8.1/11.9 & 10.6/12.2 & 5.0/15.1 & 10.9/12.0 & 4.1/7.0 & 8.3/5.2 & 10.0/8.0 & 8.9/7.9 & 3.7/8.7 & 2.0/4.8 & 0.0/0.0 \\
gte-multi-base & 25.9/29.7 & 23.3/20.3 & 12.2/24.9 & 25.3/27.5 & 18.3/27.3 & 22.8/15.8 & 24.7/20.7 & 18.0/16.9 & 19.5/23.4 & 8.1/19.9 & 0.0/0.0 \\
jina-emb-v5-small & 36.8/45.1 & 43.3/42.5 & 22.7/39.1 & 35.9/37.6 & 26.1/41.1 & 27.0/18.2 & 38.6/26.6 & 26.3/24.4 & 21.1/30.6 & 16.9/28.6 & \cellcolor{best}\textbf{42.0}/44.4 \\
jina-code-0.5b & 37.4/46.4 & 46.2/46.1 & 26.3/44.7 & 41.6/41.0 & 34.7/47.0 & 37.7/24.7 & 43.0/30.5 & 32.3/27.8 & 28.7/40.4 & 19.7/34.3 & 8.4/22.2 \\
Qwen3-0.6B & 38.2/46.1 & 37.7/37.2 & 27.5/43.8 & 34.8/33.5 & 34.1/44.7 & 30.4/19.5 & 37.4/26.2 & 28.8/25.6 & 26.0/31.3 & 12.0/23.2 & 7.1/\underline{55.6} \\
\cmidrule(lr){1-12}
\rowcolor{black!4}\multicolumn{12}{c}{\textbf{\textit{Medium Dense Retrieval Models \hspace{0.35em}(1B--2B)}}}\\
\cmidrule(lr){1-12}
gte-Qwen2-1.5B & 5.9/10.1 & 9.1/14.6 & 5.8/15.8 & 6.5/13.5 & 4.1/8.0 & 8.2/7.1 & 7.8/8.1 & 8.7/11.0 & 3.2/7.2 & 3.7/7.0 & 0.0/0.0 \\
jina-code-1.5b & 33.1/46.0 & 46.0/46.5 & 23.8/40.3 & 37.1/37.9 & 31.0/44.1 & 31.4/19.8 & 35.5/24.6 & 27.2/25.2 & 21.1/33.1 & 19.6/32.4 & 0.0/11.1 \\
F2LLM-v2-1.7B & 9.1/14.1 & 9.9/9.5 & 4.6/11.4 & 5.7/10.1 & 4.8/6.2 & 13.1/8.0 & 6.6/7.0 & 7.6/9.8 & 0.7/4.9 & 0.9/3.4 & 0.0/0.0 \\
\cmidrule(lr){1-12}
\rowcolor{black!4}\multicolumn{12}{c}{\textbf{\textit{Large Dense Retrieval Models \hspace{0.35em}($\geq$4B)}}}\\
\cmidrule(lr){1-12}
F2LLM-v2-4B & 38.5/45.1 & 45.2/41.5 & 34.8/46.5 & 40.3/37.2 & 38.5/43.8 & 35.4/20.7 & 40.6/27.3 & 28.1/25.4 & 30.0/38.3 & 21.5/33.2 & 0.0/\underline{55.6} \\
Qwen3-4B & 38.8/46.5 & 33.4/36.6 & 25.4/42.5 & 33.5/35.2 & 32.2/43.3 & 28.0/14.3 & 37.8/28.5 & 30.7/27.8 & 17.9/25.5 & 11.3/22.7 & 0.0/33.3 \\
pplx-embed-4b & 30.4/38.4 & 34.0/30.3 & 18.6/33.0 & 29.2/28.8 & 20.6/29.4 & 21.5/11.4 & 31.9/21.3 & 17.0/15.5 & 16.7/24.5 & 8.9/18.1 & 0.0/11.1 \\
C2LLM-7B & 33.7/43.2 & 46.1/44.2 & 32.5/49.4 & 40.4/37.6 & 30.8/43.9 & 35.9/19.7 & 40.8/29.5 & 28.1/27.0 & 27.3/41.7 & 14.6/31.3 & 23.6/\textbf{66.7} \\
e5-mistral-7b & 38.7/46.6 & 45.6/42.8 & 20.6/41.6 & 31.3/35.0 & 29.1/44.5 & 27.2/19.0 & 34.3/28.7 & 31.6/28.1 & 33.2/42.7 & 18.2/38.9 & 0.0/22.2 \\
SweRankEmbed-L & 42.6/47.9 & 44.3/39.1 & 24.0/39.3 & 35.0/36.0 & 34.9/42.4 & 25.9/17.2 & 33.9/28.2 & 23.5/22.0 & 23.2/28.0 & 19.4/33.2 & 0.0/44.4 \\
F2LLM-v2-8B & 36.9/44.4 & 38.2/30.4 & 21.3/29.6 & 34.3/30.8 & 32.3/29.0 & 35.8/23.0 & 36.2/26.6 & 33.0/27.5 & 29.1/29.7 & 24.2/38.2 & 14.8/22.2 \\
Qwen3-8B & 39.4/46.4 & 32.2/36.0 & 23.5/40.0 & 40.6/41.5 & 33.4/38.3 & 32.1/17.3 & 45.6/33.7 & 35.9/30.1 & 31.9/40.0 & 13.9/27.1 & 32.3/44.4 \\
\cmidrule(lr){1-12}
\rowcolor{black!4}\multicolumn{12}{c}{\textbf{\textit{In-domain SFT Dense Retrieval Models \hspace{0.35em}(0.6B--8B)}}}\\
\cmidrule(lr){1-12}
Qwen3-0.6B-SFT & 48.8/57.1 & 56.7/51.8 & 43.3/60.9 & 46.8/45.9 & 46.4/\underline{59.5} & 44.7/30.7 & 48.6/36.8 & 40.4/34.0 & 38.7/50.3 & 25.3/44.3 & 18.5/\underline{55.6} \\
Qwen3-4B-SFT & \cellcolor{second}\underline{51.6}/\textbf{63.1} & \cellcolor{second}\underline{61.6}/\underline{57.6} & \cellcolor{second}\underline{50.7}/\textbf{69.5} & \cellcolor{second}\underline{50.4}/\underline{51.6} & \cellcolor{best}\textbf{53.4}/\textbf{66.7} & \cellcolor{second}\underline{45.7}/\underline{35.7} & \cellcolor{second}\underline{50.3}/\underline{41.4} & \cellcolor{second}\underline{47.1}/\underline{40.2} & \cellcolor{second}\underline{50.3}/\textbf{68.8} & \cellcolor{second}\underline{30.2}/\underline{51.9} & \cellcolor{second}\underline{38.3}/44.4 \\
Qwen3-8B-SFT & \cellcolor{best}\textbf{53.7}/\underline{62.1} & \cellcolor{best}\textbf{64.3}/\textbf{59.1} & \cellcolor{best}\textbf{51.9}/\underline{67.0} & \cellcolor{best}\textbf{52.2}/\textbf{53.0} & \cellcolor{second}\underline{51.5}/58.7 & \cellcolor{best}\textbf{48.1}/\textbf{37.5} & \cellcolor{best}\textbf{52.6}/\textbf{42.7} & \cellcolor{best}\textbf{48.3}/\textbf{42.5} & \cellcolor{best}\textbf{53.1}/\underline{67.1} & \cellcolor{best}\textbf{30.4}/\textbf{58.5} & 23.2/\underline{55.6} \\
\bottomrule
\end{tabular}}
\end{table*}

\begin{table*}[t]
\centering
\scriptsize
\caption{Query-intent results.}
\label{tab:query-intent-results-full}
\setlength{\tabcolsep}{1.2pt}
\renewcommand{\arraystretch}{1.06}
\resizebox{\textwidth}{!}{%
\begin{tabular}{@{}l*{12}{c}@{}}
\toprule
 & \multicolumn{6}{c}{\textbf{LEVEL-2}} & \multicolumn{6}{c}{\textbf{LEVEL-3}} \\
\cmidrule(lr){2-7} \cmidrule(lr){8-13}
\textbf{Model} & \rotatebox{65}{\textbf{Bug}} & \rotatebox{65}{\textbf{Feature}} & \rotatebox{65}{\textbf{Refactor}} & \rotatebox{65}{\textbf{Question}} & \rotatebox{65}{\textbf{Docs}} & \rotatebox{65}{\textbf{Other}} & \rotatebox{65}{\textbf{Bug}} & \rotatebox{65}{\textbf{Feature}} & \rotatebox{65}{\textbf{Refactor}} & \rotatebox{65}{\textbf{Question}} & \rotatebox{65}{\textbf{Docs}} & \rotatebox{65}{\textbf{Other}} \\
\cmidrule(lr){1-13}
\rowcolor{black!4}\multicolumn{13}{c}{\textbf{\textit{Sparse Retrieval}}}\\
\cmidrule(lr){1-13}
BM25 & 11.7/35.8 & 15.3/37.6 & 20.3/33.6 & 16.2/36.1 & 7.0/36.3 & 13.3/29.9 & 17.5/24.0 & 24.2/27.4 & 25.9/27.0 & 12.5/18.4 & 20.0/23.0 & 21.1/18.1 \\
\cmidrule(lr){1-13}
\rowcolor{black!4}\multicolumn{13}{c}{\textbf{\textit{Small Dense Retrieval Models \hspace{0.35em}(<1B)}}}\\
\cmidrule(lr){1-13}
CodeRankEmbed & 8.9/28.6 & 14.3/36.6 & 23.1/36.0 & 15.4/35.0 & 9.3/30.3 & 10.3/24.4 & 21.0/24.7 & 28.7/29.6 & 36.9/29.3 & 17.9/17.9 & 28.5/27.5 & 29.6/22.3 \\
mE5-small & 8.8/31.1 & 13.8/35.0 & 24.3/36.4 & 14.8/37.9 & 19.5/23.6 & 10.0/23.1 & 18.1/24.4 & 25.1/27.4 & 35.4/30.5 & 12.7/13.6 & 17.6/18.9 & 19.0/20.0 \\
mE5-base & 2.8/13.8 & 4.7/15.7 & 7.5/11.4 & 5.6/18.6 & 1.4/15.4 & 5.2/12.4 & 5.4/9.3 & 8.2/11.6 & 11.3/10.1 & 2.3/4.7 & 9.6/9.5 & 8.3/8.7 \\
mE5-large & 8.9/31.9 & 12.0/32.5 & 19.6/30.8 & 11.7/34.2 & 21.5/26.4 & 11.3/26.4 & 15.7/22.5 & 21.4/24.1 & 26.9/24.7 & 15.4/21.3 & 19.8/17.2 & 24.5/20.3 \\
bge-large-v1.5 & 11.3/36.5 & 15.3/35.7 & 21.4/33.0 & 17.7/41.3 & 11.5/35.7 & 14.2/28.0 & 18.5/25.8 & 24.9/25.8 & 33.5/28.5 & 13.7/20.0 & 29.7/21.4 & 22.1/24.3 \\
bge-m3 & 4.0/16.5 & 4.8/15.1 & 8.3/14.8 & 5.6/20.7 & 3.1/9.7 & 4.6/14.8 & 6.7/10.8 & 9.6/11.7 & 13.2/10.8 & 6.2/7.0 & 11.7/10.3 & 11.8/10.2 \\
gte-multi-base & 10.9/32.6 & 13.7/34.9 & 20.9/30.6 & 15.7/37.3 & 8.5/28.8 & 9.8/29.2 & 19.0/23.6 & 24.1/24.9 & 30.1/24.6 & 11.2/11.3 & 21.3/22.7 & 22.1/20.0 \\
jina-emb-v5-small & 15.0/43.8 & 23.1/51.7 & 31.1/46.6 & 17.6/47.8 & 19.9/43.2 & 13.6/36.8 & 28.5/36.7 & 40.2/41.6 & 47.6/40.7 & 20.8/25.1 & 30.9/33.7 & 38.1/32.1 \\
jina-code-0.5b & 19.9/52.4 & 24.5/52.4 & 31.3/48.7 & 22.1/50.9 & 27.7/44.0 & 17.4/34.9 & 33.7/42.2 & 42.2/43.3 & 51.8/42.6 & 27.4/30.9 & 39.0/35.0 & 43.2/34.4 \\
Qwen3-0.6B & 15.0/44.6 & 19.9/47.9 & 27.5/44.9 & 14.8/43.3 & 15.4/41.3 & 9.9/33.8 & 30.8/37.7 & 37.9/38.9 & 45.4/39.4 & 27.8/29.5 & \cellcolor{second}\underline{39.2}/31.9 & 37.7/31.3 \\
\cmidrule(lr){1-13}
\rowcolor{black!4}\multicolumn{13}{c}{\textbf{\textit{Medium Dense Retrieval Models \hspace{0.35em}(1B--2B)}}}\\
\cmidrule(lr){1-13}
gte-Qwen2-1.5B & 1.8/10.3 & 4.0/17.0 & 6.2/13.6 & 2.5/14.2 & 2.5/10.1 & 3.6/10.4 & 5.6/10.5 & 8.8/14.5 & 8.0/11.5 & 9.9/9.8 & 4.6/6.5 & 4.9/11.1 \\
jina-code-1.5b & 16.0/48.8 & 20.6/50.4 & 26.2/45.0 & 17.1/48.1 & 18.1/32.7 & 13.0/35.4 & 30.5/39.5 & 39.7/42.3 & 43.4/39.5 & 27.2/28.9 & 23.7/31.3 & 39.0/30.5 \\
F2LLM-v2-1.7B & 3.8/16.8 & 6.1/19.4 & 7.2/14.1 & 6.8/24.4 & 0.0/11.1 & 4.8/16.8 & 6.5/9.7 & 9.1/11.2 & 10.4/9.4 & 6.2/3.6 & 0.0/5.4 & 8.3/8.2 \\
\cmidrule(lr){1-13}
\rowcolor{black!4}\multicolumn{13}{c}{\textbf{\textit{Large Dense Retrieval Models \hspace{0.35em}($\geq$4B)}}}\\
\cmidrule(lr){1-13}
F2LLM-v2-4B & 18.7/49.7 & 22.7/50.0 & 30.2/45.8 & 17.5/46.1 & 17.8/40.5 & 15.8/38.9 & 35.5/40.2 & 43.6/40.3 & 46.9/39.3 & 29.0/27.4 & 24.9/30.6 & 36.1/33.2 \\
Qwen3-4B & 15.0/44.8 & 20.4/48.4 & 26.1/42.3 & 17.9/44.1 & 16.8/40.7 & 11.9/34.7 & 28.9/37.0 & 36.8/39.3 & 43.2/37.2 & 19.4/28.9 & 26.7/29.4 & 32.5/30.0 \\
pplx-embed-4b & 11.7/35.6 & 15.6/38.7 & 24.0/40.9 & 12.7/40.0 & 13.7/25.5 & 8.6/22.8 & 23.3/29.2 & 30.8/30.3 & 39.2/33.7 & 17.2/22.6 & 28.3/25.5 & 30.8/23.4 \\
C2LLM-7B & 15.0/43.6 & 22.5/49.9 & 27.7/45.5 & 15.5/43.6 & 21.0/42.2 & 12.9/34.7 & 32.5/40.6 & 42.4/42.2 & 49.4/40.6 & 28.9/32.6 & 37.6/32.6 & 38.5/31.9 \\
e5-mistral-7b & 16.6/50.0 & 23.4/52.9 & 26.5/44.7 & 21.7/51.2 & 23.1/47.0 & 14.1/36.9 & 30.7/39.7 & 39.7/41.2 & 40.4/37.9 & 28.1/30.1 & 38.5/34.6 & 31.1/30.6 \\
SweRankEmbed-L & 22.1/50.9 & 25.1/50.5 & 34.1/46.8 & 20.9/46.8 & 24.9/43.9 & 15.6/36.8 & 32.7/38.2 & 39.5/38.6 & 45.5/39.0 & 23.5/23.4 & 37.4/31.1 & 32.5/28.4 \\
F2LLM-v2-8B & 16.5/44.7 & 20.5/42.5 & 28.2/37.8 & 20.4/43.0 & 23.3/35.3 & 14.7/29.2 & 30.5/33.2 & 37.8/32.4 & 38.0/29.5 & 27.4/23.6 & 35.5/29.6 & 32.7/23.7 \\
Qwen3-8B & 17.8/47.2 & 22.3/48.9 & 28.2/41.8 & 17.8/43.8 & 21.2/47.5 & 16.1/37.3 & 30.5/38.1 & 37.8/38.9 & 44.6/36.5 & 26.6/28.3 & 36.6/29.6 & 36.3/30.6 \\
\cmidrule(lr){1-13}
\rowcolor{black!4}\multicolumn{13}{c}{\textbf{\textit{In-domain SFT Dense Retrieval Models \hspace{0.35em}(0.6B--8B)}}}\\
\cmidrule(lr){1-13}
Qwen3-0.6B-SFT & 26.1/60.7 & 31.1/60.3 & 38.5/55.5 & \cellcolor{second}\underline{28.1}/55.7 & 31.0/\underline{58.9} & 23.6/48.8 & 44.7/52.4 & 53.2/51.0 & 59.7/49.2 & 44.3/48.2 & 36.0/44.2 & 44.4/41.7 \\
Qwen3-4B-SFT & \cellcolor{second}\underline{29.7}/\textbf{68.1} & \cellcolor{second}\underline{34.3}/\underline{65.4} & \cellcolor{second}\underline{41.5}/\textbf{58.4} & 27.7/\underline{61.0} & \cellcolor{second}\underline{35.7}/\textbf{60.9} & \cellcolor{second}\underline{30.4}/\underline{56.9} & \cellcolor{second}\underline{50.0}/\textbf{59.8} & \cellcolor{second}\underline{57.0}/\underline{56.6} & \cellcolor{best}\textbf{63.9}/\textbf{54.2} & \cellcolor{best}\textbf{49.3}/\textbf{52.9} & 32.1/\textbf{46.8} & \cellcolor{best}\textbf{52.1}/\underline{44.3} \\
Qwen3-8B-SFT & \cellcolor{best}\textbf{32.4}/\underline{68.1} & \cellcolor{best}\textbf{36.9}/\textbf{66.2} & \cellcolor{best}\textbf{43.1}/\underline{57.4} & \cellcolor{best}\textbf{28.9}/\textbf{63.0} & \cellcolor{best}\textbf{36.1}/57.2 & \cellcolor{best}\textbf{39.4}/\textbf{58.3} & \cellcolor{best}\textbf{51.4}/\underline{59.0} & \cellcolor{best}\textbf{59.7}/\textbf{56.7} & \cellcolor{second}\underline{62.7}/\underline{53.5} & \cellcolor{second}\underline{48.3}/\underline{52.4} & \cellcolor{best}\textbf{47.1}/\underline{46.3} & \cellcolor{second}\underline{48.8}/\textbf{44.6} \\
\bottomrule
\end{tabular}}
\end{table*}

\begin{table*}[t]
\centering
\scriptsize
\caption{Repository-difficulty results.}
\label{tab:difficulty-results-full}
\setlength{\tabcolsep}{1.2pt}
\renewcommand{\arraystretch}{1.06}
\resizebox{\textwidth}{!}{%
\begin{tabular}{@{}l*{8}{c}@{}}
\toprule
 & \multicolumn{4}{c}{\textbf{LEVEL-2}} & \multicolumn{4}{c}{\textbf{LEVEL-3}} \\
\cmidrule(lr){2-5} \cmidrule(lr){6-9}
\textbf{Model} & \rotatebox{65}{\textbf{F-S}} & \rotatebox{65}{\textbf{F-L}} & \rotatebox{65}{\textbf{D-S}} & \rotatebox{65}{\textbf{D-L}} & \rotatebox{65}{\textbf{F-S}} & \rotatebox{65}{\textbf{F-L}} & \rotatebox{65}{\textbf{D-S}} & \rotatebox{65}{\textbf{D-L}} \\
\cmidrule(lr){1-9}
\rowcolor{black!4}\multicolumn{9}{c}{\textbf{\textit{Sparse Retrieval}}}\\
\cmidrule(lr){1-9}
BM25 & 12.2/50.0 & 12.4/34.6 & 16.4/42.3 & 14.4/30.9 & -- & 5.0/32.9 & 15.2/43.8 & 20.9/23.4 \\
\cmidrule(lr){1-9}
\rowcolor{black!4}\multicolumn{9}{c}{\textbf{\textit{Small Dense Retrieval Models \hspace{0.35em}(<1B)}}}\\
\cmidrule(lr){1-9}
CodeRankEmbed & 13.4/43.7 & 9.5/28.8 & 14.5/39.7 & 12.6/28.4 & -- & 12.4/11.1 & 24.4/44.2 & 24.7/25.0 \\
mE5-small & 10.3/45.0 & 9.6/30.2 & 15.0/39.8 & 13.3/29.0 & -- & 3.2/11.1 & 19.2/43.5 & 21.7/23.9 \\
mE5-base & 4.3/25.4 & 2.6/11.1 & 6.6/26.9 & 4.4/11.6 & -- & 0.0/5.6 & 6.2/24.8 & 6.8/8.7 \\
mE5-large & 9.9/43.9 & 9.6/31.3 & 12.7/37.2 & 11.6/26.2 & -- & 17.1/36.4 & 15.5/39.4 & 18.8/21.5 \\
bge-large-v1.5 & 13.5/51.8 & 12.4/35.2 & 16.8/43.4 & 13.1/28.2 & -- & 0.0/2.8 & 21.0/48.5 & 21.7/23.7 \\
bge-m3 & 4.7/25.8 & 3.5/13.5 & 6.9/26.2 & 5.3/12.9 & -- & 9.4/11.1 & 9.9/28.4 & 8.0/9.4 \\
gte-multi-base & 12.8/46.6 & 11.7/31.5 & 15.4/41.8 & 12.1/27.8 & -- & 14.7/41.7 & 22.3/43.3 & 21.3/22.1 \\
jina-emb-v5-small & 17.9/56.9 & 17.2/45.5 & 21.6/50.9 & 19.5/41.9 & -- & 6.4/30.2 & 31.1/54.6 & 34.0/36.9 \\
jina-code-0.5b & 21.8/62.1 & 21.1/53.1 & 21.8/51.3 & 23.3/45.5 & -- & \cellcolor{best}\textbf{37.3}/\underline{71.1} & 34.3/57.4 & 38.1/40.8 \\
Qwen3-0.6B & 17.4/54.3 & 16.1/45.6 & 18.6/46.6 & 17.8/40.9 & -- & 12.8/47.1 & 35.2/52.3 & 34.1/36.6 \\
\cmidrule(lr){1-9}
\rowcolor{black!4}\multicolumn{9}{c}{\textbf{\textit{Medium Dense Retrieval Models \hspace{0.35em}(1B--2B)}}}\\
\cmidrule(lr){1-9}
gte-Qwen2-1.5B & 2.9/21.5 & 1.7/8.6 & 6.5/25.1 & 3.4/12.1 & -- & 0.0/1.0 & 8.9/26.1 & 6.7/10.6 \\
jina-code-1.5b & 16.0/58.0 & 17.6/49.8 & 19.2/48.9 & 18.9/43.0 & -- & 16.7/63.8 & 32.0/55.9 & 34.8/38.7 \\
F2LLM-v2-1.7B & 6.3/30.7 & 3.7/14.4 & 9.5/33.2 & 4.4/13.1 & -- & 0.0/8.3 & 9.1/24.6 & 7.5/8.8 \\
\cmidrule(lr){1-9}
\rowcolor{black!4}\multicolumn{9}{c}{\textbf{\textit{Large Dense Retrieval Models \hspace{0.35em}($\geq$4B)}}}\\
\cmidrule(lr){1-9}
F2LLM-v2-4B & 19.1/58.0 & 20.1/50.9 & 21.4/50.1 & 21.2/42.6 & -- & \cellcolor{second}\underline{36.7}/62.6 & 37.6/51.8 & 39.0/38.8 \\
Qwen3-4B & 16.9/55.2 & 16.0/45.3 & 19.9/48.8 & 18.6/41.2 & -- & 10.1/38.0 & 33.2/55.1 & 32.4/36.0 \\
pplx-embed-4b & 13.3/45.3 & 12.7/36.1 & 14.6/42.1 & 14.4/32.3 & -- & 5.7/26.0 & 23.4/44.6 & 27.2/28.3 \\
C2LLM-7B & 18.4/53.0 & 16.6/46.0 & 20.6/47.4 & 19.4/41.0 & -- & 8.2/22.5 & 37.0/56.7 & 37.0/39.5 \\
e5-mistral-7b & 19.8/60.6 & 18.4/51.3 & 21.4/52.1 & 19.9/44.1 & -- & 10.2/59.0 & 33.8/61.1 & 34.6/37.9 \\
SweRankEmbed-L & 24.1/61.7 & 23.1/50.8 & 22.5/51.7 & 24.0/43.8 & -- & 24.6/64.1 & 32.6/54.8 & 36.1/36.4 \\
F2LLM-v2-8B & 14.2/49.1 & 17.9/44.5 & 21.3/47.3 & 20.2/37.9 & -- & 26.0/26.7 & 25.6/38.3 & 34.3/32.0 \\
Qwen3-8B & 18.8/55.9 & 19.0/48.2 & 21.8/48.3 & 20.6/41.7 & -- & 12.8/33.5 & 29.6/50.9 & 34.4/36.9 \\
\cmidrule(lr){1-9}
\rowcolor{black!4}\multicolumn{9}{c}{\textbf{\textit{In-domain SFT Dense Retrieval Models \hspace{0.35em}(0.6B--8B)}}}\\
\cmidrule(lr){1-9}
Qwen3-0.6B-SFT & 28.2/67.9 & 28.0/62.1 & 26.0/55.4 & 29.2/54.2 & -- & 30.0/\textbf{72.1} & 45.5/68.7 & 48.8/49.8 \\
Qwen3-4B-SFT & \cellcolor{second}\underline{30.9}/\textbf{74.7} & \cellcolor{second}\underline{31.8}/\underline{69.2} & \cellcolor{second}\underline{29.1}/\textbf{61.2} & \cellcolor{second}\underline{32.3}/\underline{59.4} & -- & 29.7/69.2 & \cellcolor{best}\textbf{50.4}/\textbf{75.7} & \cellcolor{second}\underline{53.5}/\underline{56.3} \\
Qwen3-8B-SFT & \cellcolor{best}\textbf{34.0}/\underline{73.2} & \cellcolor{best}\textbf{35.0}/\textbf{69.7} & \cellcolor{best}\textbf{32.7}/\underline{60.5} & \cellcolor{best}\textbf{33.7}/\textbf{60.4} & -- & 16.8/53.3 & \cellcolor{second}\underline{49.5}/\underline{69.5} & \cellcolor{best}\textbf{55.4}/\textbf{56.5} \\
\bottomrule
\end{tabular}}
\end{table*}

\section{Prompt Templates}
\label{sec:prompt-appendix}
\label{app:prompt-templates}

The following boxes show the prompt templates used by the data construction
pipeline. For compactness, each box merges the system message and user-message
template. Variables in angle brackets are filled by the pipeline.
Table~\ref{tab:prompt-query-rewrite} gives the rewrite prompt used to turn
retained PR or issue text into developer-style queries while keeping failure
messages and technical clues faithful to the original request.
Table~\ref{tab:prompt-query-filter} shows the separate filtering prompt, which
removes answer-leaking, underspecified, or nearly empty queries before benchmark
construction.
Table~\ref{tab:prompt-relevance-judge} lists the relevance-judging prompt used
to turn trajectory-derived candidate chunks into LEVEL-3 labels.

\section{Usage of AI Assistant}
For manuscript preparation, we used LLM-based assistance for language
polishing and grammar refinement.

\begin{table*}[p]
\centering
\caption{Query rewriting prompt.}
\label{tab:prompt-query-rewrite}
\begin{tcblisting}{promptbox,title={Query rewriting prompt}}
[System prompt]
You are an expert at analyzing and rewriting GitHub issue texts for a code-retrieval benchmark.
You have two tasks: (1) classify the issue, (2) rewrite it as a realistic vibe-coding query a developer would type into an AI coding assistant.
Output only a valid JSON object and nothing else.

[User prompt template]
Analyze the following GitHub issue and perform TWO tasks:

## Task 1: Classify the issue on three dimensions

- **description_style**: "natural_language" if the issue is primarily prose description; "error_traceback" if it is primarily structured around a stack trace, error output, or log messages.
- **intent_type**: Choose the single best fit from: "bug_report", "feature_request", "question", "refactoring", "documentation", "other".
- **content_type**: "text_heavy" if the content is mostly prose with little or no code/paths; "code_heavy" if the dominant content is code blocks or snippets; "file_path_heavy" if the dominant non-prose content is file paths or module names; "mixed" otherwise.

## Task 2: Rewrite the issue as a vibe-coding developer query

Imagine a developer encounters this exact problem while coding and pastes it into an AI coding assistant. Rewrite the issue as the raw, unpolished query they would type.

### What vibe-coding queries look like
- The developer describes symptoms, not root causes: "my X is broken when I do Y" not "the internal state machine fails to transition".
- They paste error messages and stack traces directly without editing them.
- They may include a short code snippet showing what they tried.
- Tone is casual and direct: "why does this crash?", "how do I fix this?", "getting a weird error".
- They do NOT write structured reports: no headers, no "Expected behavior:" labels, no checklists.

### Rewriting rules

Preserve verbatim:
- All error messages, exception types, and stack traces.
- Code snippets that directly show the failing code or minimal reproducer.

Rewrite or compress:
- Translate formal issue prose into casual first-person phrasing.
- Remove GitHub issue template noise, such as HTML comments, section headers, checkboxes, and version tables.
- Drop unrelated environment details unless they are clearly relevant.
- Remove duplicate information.

Per description_style guidance:
- If `error_traceback`: lead with one casual sentence describing what the user was doing, paste the full traceback/error output as-is, then end with a short question.
- If `natural_language`: rewrite as a concise first-person question of about 50--200 words. Keep key technical terms, library names, and short code snippets that illustrate the problem.

Length:
- Target 50--300 words for the rewritten query.
- Never reproduce the entire issue verbatim.
- Never reduce to a single vague sentence that loses technical specifics.

Language: Keep the same language as the original issue.
Faithfulness: Do NOT add information, suggestions, or context not present in the original issue.

Output format:
{
  "description_style": "<natural_language|error_traceback>",
  "intent_type": "<bug_report|feature_request|question|refactoring|documentation|other>",
  "content_type": "<text_heavy|code_heavy|file_path_heavy|mixed>",
  "rewritten_query": "<the rewritten vibe-coding query>",
  "reason": "<one sentence explaining your classification choices, under 40 words>"
}

Rules:
- Output ONLY the JSON object, no markdown fences, no extra text.
- All field values must exactly match the allowed values above.

<issue_text>
<raw_issue_text>
</issue_text>
\end{tcblisting}
\end{table*}

\begin{table*}[p]
\centering
\caption{Query filtering prompt.}
\label{tab:prompt-query-filter}
\begin{tcblisting}{promptbox,title={Query filtering prompt}}
[System prompt]
You are a query quality judge for a code retrieval benchmark. Each query is a GitHub PR/Issue description used as a search query to find relevant code files.

Classify the query into one of three categories:

1. FILTER_MEANINGLESS: The query is meaningless, underspecified, non-actionable,
or has no substantive problem content.
Examples:
- Only contains an issue/PR number, such as "Fix #1234" or "Closes #567".
- Contains only boilerplate template text with no filled content.
- Is extremely short with no actionable information, such as "update", "fix", or "bug".
- Contains only links or references without describing the problem.
- Describes a broad maintenance task without symptoms, expected behavior, error
messages, APIs, or repository-specific clues.

2. FILTER_ANSWER_LEAK: The query already reveals the solution, making code retrieval trivial.
Examples:
- Explicitly mentions which file(s) to modify.
- Describes exact code changes or diffs.
- Provides the fix implementation in the description.
- Names specific functions/methods/classes to change AND describes the exact change.
Note: Simply mentioning a module name or function as part of describing a problem is NOT an answer leak. Only classify as leak when the query tells you where and how to fix it.
File paths from tracebacks, logs, or error messages are also not answer leaks
unless the query instructs the developer to edit those files or gives the patch recipe.

3. KEEP: The query is a valid, meaningful problem description with enough
issue-specific signal for retrieval and does not reveal the answer.

Respond with ONLY a JSON object in this exact format:
{"verdict": "FILTER_MEANINGLESS" | "FILTER_ANSWER_LEAK" | "KEEP", "reason": "<brief reason in English>"}

[User prompt template]
Query ID: <query_id>
Query text:
<query_text>
\end{tcblisting}
\end{table*}

\begin{table*}[p]
\centering
\caption{LLM relevance judge prompt.}
\label{tab:prompt-relevance-judge}
\begin{tcblisting}{promptbox,title={LLM relevance judge prompt}}
[System prompt]
You are a strict code-retrieval relevance judge for software issue resolution.

Your task is to decide how useful a trajectory-derived candidate code chunk is for understanding, localizing, or fixing the issue.
Return a single JSON object and nothing else.

[User prompt template]
Evaluate the relevance of the candidate code chunk for the issue below.

Issue instance: <instance_id>
Query id: <query_id>

Relevance labels:
- high: the chunk is directly about the buggy logic, the likely fix location, a closely related serializer/parser/data-flow path, or a key failing test.
- partial: the chunk is adjacent context that would help understand or localize the fix, but is not itself a strong candidate for the main fix.
- irrelevant: the chunk is unrelated, too generic, or unlikely to help understand/localize/fix the issue.

Output schema:
{
  "label": "high" | "partial" | "irrelevant",
  "score": 2 | 1 | 0,
  "confidence": 0.0 to 1.0,
  "reason": "short justification"
}

Rules:
- `score` must match `label` exactly: high=2, partial=1, irrelevant=0.
- Keep `reason` under 60 words.
- Use only the issue text and chunk content below.
- A chunk can be relevant even if it is not an edit location, as long as it helps
understand or localize the issue.
- Do not explain the schema.
- Do not use markdown fences.

<truncation_note>

<issue_text>
<raw_issue_text>
</issue_text>

<chunk_metadata>
file_path: <file_path>
start_line: <start_line>
end_line: <end_line>
language: <language>
chunk_index: <chunk_index>
</chunk_metadata>

<code_chunk>
<code_chunk_text>
</code_chunk>
\end{tcblisting}
\end{table*}

\end{document}